\RequirePackage{lineno}

\documentclass[aps,preprint,showpacs,superscriptaddress,groupedaddress,amsmath]{revtex4}

\usepackage{amsmath,amsfonts,graphicx,bm,amssymb,array}
\usepackage[usenames]{color}
\usepackage{array}
\usepackage{float}
\usepackage{sidecap}
\usepackage{graphicx}
\usepackage{epstopdf}
\usepackage{color}
\usepackage{longtable}
\usepackage{subfigure}
\usepackage{longtable}
\usepackage{multirow} 

\def\etmiss{E\!\!\!\!\slash_{T}}

\begin{document}

\title{Searches for resonances in the tb and tc final states at the high-luminosity LHC}

\author{Elizabeth Drueke} \affiliation{Department of Physics and Astronomy, Michigan State University, East Lansing MI 48824, USA}
\author{Brad Schoenrock} \affiliation{Department of Physics and Astronomy, Michigan State University, East Lansing MI 48824, USA}
\author{Barbara Alvarez Gonzalez} \affiliation{Department of Physics and Astronomy, Michigan State University, East Lansing MI 48824, USA}
\author{Reinhard Schwienhorst} \affiliation{Department of Physics and Astronomy, Michigan State University, East Lansing MI 48824, USA}

\date{\today}

\begin{abstract}
We study resonances decaying to one top quark and one additional quark ($b$ or $c$) at 
the low-luminosity and high-luminosity 14~TeV LHC and at a future 33~TeV hadron collider in
the context of Snowmass 2013. A heavy $W'$~boson that preferentially couples to quarks can
be found through its decay to $tb$. A Kaluza-Klein~gluon might have a significant branching
ratio to $tc$.
The final state in these searches has a lepton and neutrino from a W boson decay
plus two jets, at least one of which is $b$-tagged. We give expected limits as a 
function of $W'$~boson and $KKg$~masses for different collider energy and integrated luminosity
options.
\end{abstract}

\pacs{14.65.Jk, 14.65.Ha, 12.60.-i, 14.80.Rt} 
\maketitle

\modulolinenumbers[1]
\linenumbers


\section{Introduction}
\label{sec:intro}
~
The Large Hadron Collider (LHC) is the highest-energy particle accelerator ever built,
probing physics at the TeV scale. The Higgs boson discovery~\cite{Aad:2012tfa,Chatrchyan:2012ufa}
was the first, but more discoveries are likely as the LHC covers the energy range
where new physics is expected. Searches for high-mass resonances take advantage of the
high center-of-mass (CM) energy of the LHC and have been performed in many final states. The $tb$ mode,
where $tb$ corresponds to $t\bar{b}$ and $\bar{t} b$, is particularly sensitive to a heavy partner of 
the $W$~boson, i.e. a $W'$~boson coupling primarily to quarks and the third
generation~\cite{Tait:2000sh}. Such a $W'$~boson appears in new physics models that have 
additional symmetries, such as universal extra dimensions~\cite{Datta:2000gm} or little Higgs
models~\cite{Perelstein:2005ka}.

\begin{figure}[h]
  \centering
  \includegraphics[width=0.48\textwidth]{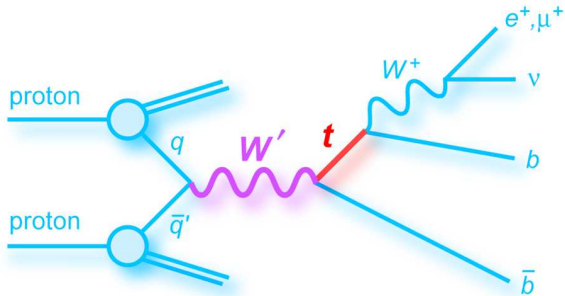}
  \caption{Feynman diagram for the production of a $W'$~boson with decay to lepton+jets.}
  \label{fig:WP_feynman}
\end{figure}

Searches for a $W'$~boson have been performed by ATLAS~\cite{Aad:2012ej, WprimeCONF} and
CMS~\cite{CMSWprime}, with $W'$~mass limits approaching 2~TeV for different $W'$~couplings.
The Feynman diagram for
$W'$~production is shown in Fig.~\ref{fig:WP_feynman}. We focus on $W'$~bosons with purely
right-handed couplings, the sensitivity to left-handed or mixed couplings is similar.

\begin{figure}[h]
  \centering
  \includegraphics[width=0.48\textwidth]{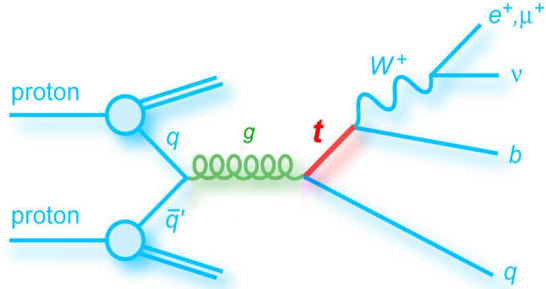}
  \caption{Feynman diagram for the production of a $KKg$ with decay to lepton+jets.}
  \label{fig:KKg_feynman}
\end{figure}

Another possibility that has not been explored experimentally yet is the flavor-changing
neutral current (FCNC) decay of a Kaluza-Klein gluon
($KKg$)~\cite{Aquino:2006vp}. We study a $KKg$ that has a non-negligible branching ratio to
$tc$, where $tc$ corresponds to $t\bar{c}$ and $\bar{t} c$~\cite{Delaunay:2010dw}. 
The Feynman diagram for $KKg$ production is shown in Fig.~\ref{fig:KKg_feynman}. It is likely
that such a particle will be discovered first in its dominant $t\bar{t}$ decay mode, and we
present the 95\% confidence level (C.L.) limit as a function of the $KKg$ mass for the
process $pp\rightarrow KKg\rightarrow tc$, at a cross-section corresponding to a 40$\%$ of
the $KKg \rightarrow t\bar{t}$ cross-section~\cite{KKgttbar}, which is evaluated at 
next-to-leading order (NLO) in QCD~\cite{NLOKKg} by scaling the leading order (LO) cross
section by a factor 1.3.

In this paper we explore the sensitivity of 14~TeV and 33~TeV proton-proton colliders
to $W'$~bosons and $KKg$ in three different
scenarios under investigation for the Snowmass 2013 process: 300~fb$^{-1}$ at 14~TeV
with an average pileup of 50~events, 3000~fb$^{-1}$ at 14~TeV
with an average pileup of 140~events, and 3000~fb$^{-1}$ at 33~TeV with an average
pileup of 140~events. For each scenario the
Snowmass LHC detector models are used~\cite{Anderson:2013kda}.

This paper is organized as follows:
Section~\ref{sec:models} introduces the $W'$ and $KKg$ models and their parameters.
Section~\ref{sec:analysis} explains the event selection. 
Section~\ref{sec:selection} present the analysis strategy.
Section~\ref{sec:results} describes the limit setting procedure and 
shows the results and, finally, Section~\ref{sec:conclusions} gives our conclusions.

%
\section{Signal and Background Modeling}
\label{sec:models}
The $W'$ and $KKg$ events are generated with Madgraph5~\cite{Alwall:2011uj} and passed through 
Pythia8~\cite{pythia} for showering and Delphes3~\cite{delphes3} for detector modeling, using
the Snowmass settings~\cite{Anderson:2013kda}.

At each mass point, $W'$ and $KKg$ events are generated with equal amounts of top and antitop
events, 50,000 each. Only $W$~boson decays to an electron or a muon are included.

In addition to these signal samples, background samples generated for Snowmass are
used~\cite{Avetisyan:2013onh}.
Those samples are diboson, $W/\gamma/Z$+Jets, $t\bar{t}$, $Wt$, $t$-channel, and $s$-channel single
top. 
Each background is normalized according to the corresponding sample cross-section. The 
$W'$~signal samples are normalized to NLO cross-sections~\cite{Sullivan} as shown in
Table~\ref{tab:wpxsec}. The $KKg \rightarrow t\bar{c}$ samples are normalized to 40\% of the
corresponding $KKg \rightarrow t\bar{t}$ cross-section at NLO as shown in
Table~\ref{tab:kkgxsec}. 

\begin{table}[h]
  \begin{center}
    \renewcommand{\arraystretch}{1.4}                      
    \begin{tabular}{lcccc}
      \hline
      \multirow{2}{*}{$W'$ Mass [TeV]} & \multicolumn{2}{c}{14~TeV} & \multicolumn{2}{c}{33~TeV} \\
                                       & $\sigma_{t\bar{b}}$ [pb] & $\sigma_{\bar{t}b}$ [pb] & $\sigma_{t\bar{b}}$ [pb] & $\sigma_{\bar{t}b}$ [pb]\\
      \hline
      1~TeV  & 9.97E+00 & 4.68E+00 & 3.84E+01 & 2.29E+01 \\
      2~TeV  & 5.14E-01 & 1.86E-01 & 3.29E+00 & 1.62E+00 \\
      3~TeV  & 5.55E-02 & 1.81E-02 & 6.65E-01 & 2.81E-01 \\
      4~TeV  & 7.72E-03 & 2.62E-03 & 1.88E-01 & 7.21E-02 \\
      5~TeV  & 1.37E-03 & 5.35E-04 & 6.37E-02 & 2.26E-02 \\
      6~TeV  & 3.71E-04 & 1.61E-04 & 2.40E-02 & 8.09E-03 \\
      7~TeV  & 1.57E-04 & 6.99E-05 & 9.53E-03 & 3.18E-03 \\
      8~TeV  &          &          & 4.00E-03 & 1.35E-03 \\
      9~TeV  &          &          & 1.76E-03 & 6.09E-04 \\
      10~TeV &          &          & 8.12E-04 & 2.97E-04\\
      11~TeV &          &          & 3.95E-04 & 1.55E-04\\
      12~TeV &          &          & 2.08E-04 & 8.71E-05\\
      \hline
    \end{tabular}
    \caption{$W'$~production cross-section at NLO at 14~TeV and 33~TeV~\cite{Sullivan}.}
    \label{tab:wpxsec}
  \end{center}
\end{table}
\begin{table}[h]
  \begin{center}                                                                         
    \renewcommand{\arraystretch}{1.4}                                                      
    \begin{tabular}{lcc}                                                        
      \hline                                                                                 
      $KKg$ Mass [TeV] & 14~TeV & 33~TeV \\           
      \hline                                                                                 
      1~TeV & 9.83E+00 & 4.43E+01 \\ 
      2~TeV & 5.20E-01 & 3.55E+00 \\      
      3~TeV & 6.55E-02 & 6.97E-01 \\ 
      4~TeV & 1.28E-02 & 1.99E-01 \\
      5~TeV & 3.90E-03 & 6.92E-02 \\        
      6~TeV & 1.42E-03 & 2.77E-02 \\ 
      7~TeV & 6.60E-04 & 1.24E-02 \\
      8~TeV &  & 5.98E-03 \\
      9~TeV &  & 3.11E-03 \\
      10~TeV &  & 1.72E-03 \\
      11~TeV &  & 1.03E-03 \\
      12~TeV &  & 8.06E-04 \\
      \hline                                                                                 
    \end{tabular}                                                                          
    \caption{$KKg \rightarrow tc$ production cross-section at NLO at 14~TeV and 33~TeV.}
    \label{tab:kkgxsec}
  \end{center}
\end{table}

%
\section{Analysis}
\label{sec:analysis}

Signal and background events are required to pass the following basic selection cuts
for both $W'$ and $KKg$. Signal-specific cuts are applied in a second step described in
Section~\ref{sec:selection}.

\begin{eqnarray}
\textrm{Exactly one lepton (electron or muon) with } \qquad p_{T}^{\ell}&\geq& 25\,{\rm GeV}, \qquad
\left|\eta_{\ell}\right|\leq 2.5, \nonumber \\
\textrm{At least two jets with } \qquad p_T^j&\geq& 25\,{\rm GeV}, \qquad 
\left|\eta_{j}\right|\leq 2.5, \nonumber \\
\textrm{At least one b-tagged jet with} \qquad p_T^j&\geq& 25\,{\rm GeV}, \qquad 
\left|\eta_{j}\right|\leq 2.5, \nonumber \\
\textrm{Missing energy} \qquad ~\etmiss &>& 25~{\rm GeV} \nonumber \\
\label{eq:basiccut}
\end{eqnarray}
where $p_{T}^{\ell}$ and $\eta_{\ell}$ correspond to the transverse momentum and pseudorapidity of
the lepton, and $p_T^j$ and $\eta_{j}$ are the transverse momentum and pseudorapidity of each jet.
Jets are reconstructed using the Cambridge-Aachen (CA) jet clustering algorithm~\cite{CAjets}
with a radius parameter of $R=0.8$, ideal for the high-$p_T$ jets and top quark decays associated
with our signal and backgrounds.

Information about the flavor content of jets is obtained by looking for a match within $\Delta R=1.0$
with a $b$-tagged anti-kt jet~\cite{CAC-0801} with radius parameter $R=0.4$. 

All samples are weighted to cross-section times integrated luminosity, including event-specific
weights:
\begin{eqnarray}
\rm \frac{Luminosity [fb^{-1}] \cdot Cross~section [fb]}{Number~of~generated~events} \cdot Event~weight\, .
\end{eqnarray}

Here, the cross-section corresponds to the signal and background cross-sections described
in~\ref{sec:models}.
The $W'$~and $KKg$~cross-sections from 
Tables~\ref{tab:wpxsec} and ~\ref{tab:kkgxsec} are multiplied by an additional factor of
$\frac{21}{100}$ to account for the $W$~boson decay branching ratios to $e$ or $\mu$.

Tables~\ref{tab:preselectionwp} and~\ref{tab:preselectionkkg20} give the
expected event yields for $W'$ and $KKg$ and the backgrounds, respectively,
for each of the scenarios under consideration. The signal-to-background ratio is shown in
Table~\ref{tab:sigbacwpbasic} for the $W'$ analysis and in Table~\ref{tab:sigbackkgbasic} for the $KKg$ analysis.
The dominant background for 300~$fb^{-1}$ at 14~TeV is $t\overline{t}$ production, followed
by $W$/$\gamma$/$Z$+jets. For the 3000~$fb^{-1}$ scenarios at both 14~TeV and 33~TeV, 
$W$/$\gamma$/$Z$+jets is the dominant background followed by $t\overline{t}$.

\begin{table}[H]
\begin{center}
\renewcommand{\arraystretch}{1.4}
\begin{tabular}{lccc}
\hline
\multirow{2}{*}{Preselection} & \multicolumn{2}{c}{14~TeV}  & 33~TeV  \\ 
 &  300~$fb^{-1}$&  3000~$fb^{-1}$  & 3000~$fb^{-1}$  \\ 
\hline
Diboson & 1.20E+05 & 6.51E+06 & 3.75E+07 \\
$W$/$\gamma$/$Z$+Jets & 2.39E+06 & 5.80E+08 & 5.39E+09 \\
$t\overline{t}$ & 4.24E+06 & 2.28E+08 & 1.59E+09 \\
$Wt$ single top & 3.30E+05 & 1.75E+07 & 1.14E+08 \\
$t/s$-channel single top & 3.31E+05 & 2.36E+07 & 1.17E+08 \\
$W'$ (1~TeV) & 3.43E+05 & 4.51E+06 & 1.73E+07 \\ 
$W'$ (2~TeV) & 1.66E+04 & 1.76E+05 & 1.14E+06 \\ 
$W'$ (3~TeV) & 1.24E+03 & 1.27E+04 & 1.53E+05 \\ 
$W'$ (4~TeV) & 1.25E+02 & 1.30E+03 & 2.95E+04 \\ 
$W'$ (5~TeV) & 1.79E+01 & 1.88E+02 & 7.45E+03 \\ 
$W'$ (6~TeV) & 4.88E+00 & 5.50E+01 & 2.26E+03 \\ 
$W'$ (7~TeV) & 2.63E+00 & 3.41E+01 & 7.61E+02 \\
$W'$ (8~TeV) &  &  & 2.87E+02 \\
$W'$ (9~TeV) &  &  & 1.15E+02 \\
$W'$ (10~TeV) &  &  & 5.25E+01 \\
$W'$ (11~TeV) &  &  & 2.47E+01 \\
$W'$ (12~TeV) &  &  & 1.36E+01 \\
\hline
\end{tabular}
\caption{Signal and background event yields after the preselection
for various pileup configurations, luminosities, and energy levels for the $W'$ analysis.} 
\label{tab:preselectionwp}
\end{center}
\end{table}

\begin{table}[H]
\begin{center}
\renewcommand{\arraystretch}{1.4}
\begin{tabular}{lccc}
\hline
\multirow{2}{*}{Preselection} & \multicolumn{2}{c}{14~TeV}  & 33~TeV  \\ 
 &  300~$fb^{-1}$&  3000~$fb^{-1}$  & 3000~$fb^{-1}$  \\ 
\hline
Diboson & 1.20E+05 & 6.51E+06 & 3.75E+07 \\
$W$/$\gamma$/$Z$+Jets & 2.39E+06 & 5.80E+08 & 5.39E+09 \\
$t\overline{t}$ & 4.24E+06 & 2.28E+08 & 1.59E+09 \\
$Wt$ single top & 3.30E+05 & 1.75E+07 & 1.14E+08 \\
$t/s$-channel single top & 3.31E+05 & 2.36E+07 & 1.17E+08 \\
$KKg$ (1~TeV) & 1.60E+05 & 2.28E+06 & 9.38E+06 \\
$KKg$ (2~TeV) & 8.85E+03 & 1.03E+05 & 6.34E+05 \\
$KKg$ (3~TeV) & 8.77E+02 & 1.03E+04 & 9.45E+04 \\
$KKg$ (4~TeV) & 1.52E+02 & 1.91E+03 & 2.17E+04 \\
$KKg$ (5~TeV) & 4.65E+01 & 6.20E+02 & 6.90E+03 \\
$KKg$ (6~TeV) & 1.77E+01 & 2.47E+02 & 2.71E+03 \\
$KKg$ (7~TeV) & 8.44E+00 & 1.21E+02 & 1.26E+03 \\
$KKg$ (8~TeV) &  &  & 6.57E+02 \\
$KKg$ (9~TeV) &  &  & 3.61E+02 \\
$KKg$ (10~TeV) &  &  & 2.09E+02 \\
$KKg$ (11~TeV) &  &  & 1.28E+02 \\
$KKg$ (12~TeV) &  &  & 8.36E+01 \\
\hline
\end{tabular}
\caption{Signal and background event yields after the preselection
for various pileup configurations, luminosities, and energy levels for the $KKg$ analysis.} 

\label{tab:preselectionkkg20}
\end{center}
\end{table}

\begin{table}[H]
\begin{center}
\renewcommand{\arraystretch}{1.4}
\begin{tabular}{lccc}
\hline
\multirow{2}{*}{Preselection} & \multicolumn{2}{c}{14~TeV}  & 33~TeV  \\ 
 &  300~$fb^{-1}$&  3000~$fb^{-1}$  & 3000~$fb^{-1}$  \\ 
\hline
$W'$ (1~TeV) & 4.63E-02 & 5.25E-03 & 6.22E-05 \\
$W'$ (2~TeV) & 2.24E-03 & 2.04E-04 & 4.20E-06 \\
$W'$ (3~TeV) & 1.67E-04 & 1.48E-05 & 6.27E-07 \\
$W'$ (4~TeV) & 1.69E-05 & 1.51E-06 & 1.44E-07 \\
$W'$ (5~TeV) & 2.42E-06 & 2.19E-07 & 4.58E-08 \\
$W'$ (6~TeV) & 6.59E-07 & 6.39E-08 & 1.80E-08 \\
$W'$ (7~TeV) & 3.56E-07 & 3.96E-08 & 8.37E-09 \\
$W'$ (8~TeV) &  &  & 4.36E-09 \\
$W'$ (9~TeV) &  &  & 2.40E-09 \\
$W'$ (10~TeV) &  &  & 1.39E-09 \\
$W'$ (11~TeV) &  &  & 8.51E-10 \\
$W'$ (12~TeV) &  &  & 5.55E-10 \\
\hline
\end{tabular}
\caption{Ratio of signal over background event yields after the preselection for various
$W'$~masses.} 
\label{tab:sigbacwpbasic}
\end{center}
\end{table}

\begin{table}[H]
\begin{center}
\renewcommand{\arraystretch}{1.4}
\begin{tabular}{lccc}
\hline
\multirow{2}{*}{Preselection} & \multicolumn{2}{c}{14~TeV}  & 33~TeV  \\ 
 &  300~$fb^{-1}$&  3000~$fb^{-1}$  & 3000~$fb^{-1}$  \\ 
\hline
$KKg$ (1~TeV) & 2.16E-02 & 2.67E-03 & 1.29E-03 \\
$KKg$ (2~TeV) & 1.19E-03 & 1.20E-04 & 8.74E-05 \\
$KKg$ (3~TeV) & 1.18E-04 & 1.20E-05 & 1.30E-05 \\
$KKg$ (4~TeV) & 2.05E-05 & 2.24E-06 & 3.00E-06 \\
$KKg$ (5~TeV) & 6.28E-06 & 7.25E-07 & 9.52E-07 \\
$KKg$ (6~TeV) & 2.39E-06 & 2.89E-07 & 3.74E-07 \\
$KKg$ (7~TeV) & 1.14E-06 & 1.42E-07 & 1.74E-07 \\
$KKg$ (8~TeV) &  &  & 9.06E-08 \\
$KKg$ (9~TeV) &  &  & 4.98E-08 \\
$KKg$ (10~TeV) &  &  & 2.89E-08 \\
$KKg$ (11~TeV) &  &  & 1.77E-08 \\
$KKg$ (12~TeV) &  &  & 1.15E-08 \\
\hline
\end{tabular}
\caption{Ratio of signal over background event yields after the preselection
for various $KKg$~masses.}
\label{tab:sigbackkgbasic}
\end{center}
\end{table}

\begin{figure}[H]
  \centering
  \subfigure[]{
    \includegraphics[width=0.37\textwidth]{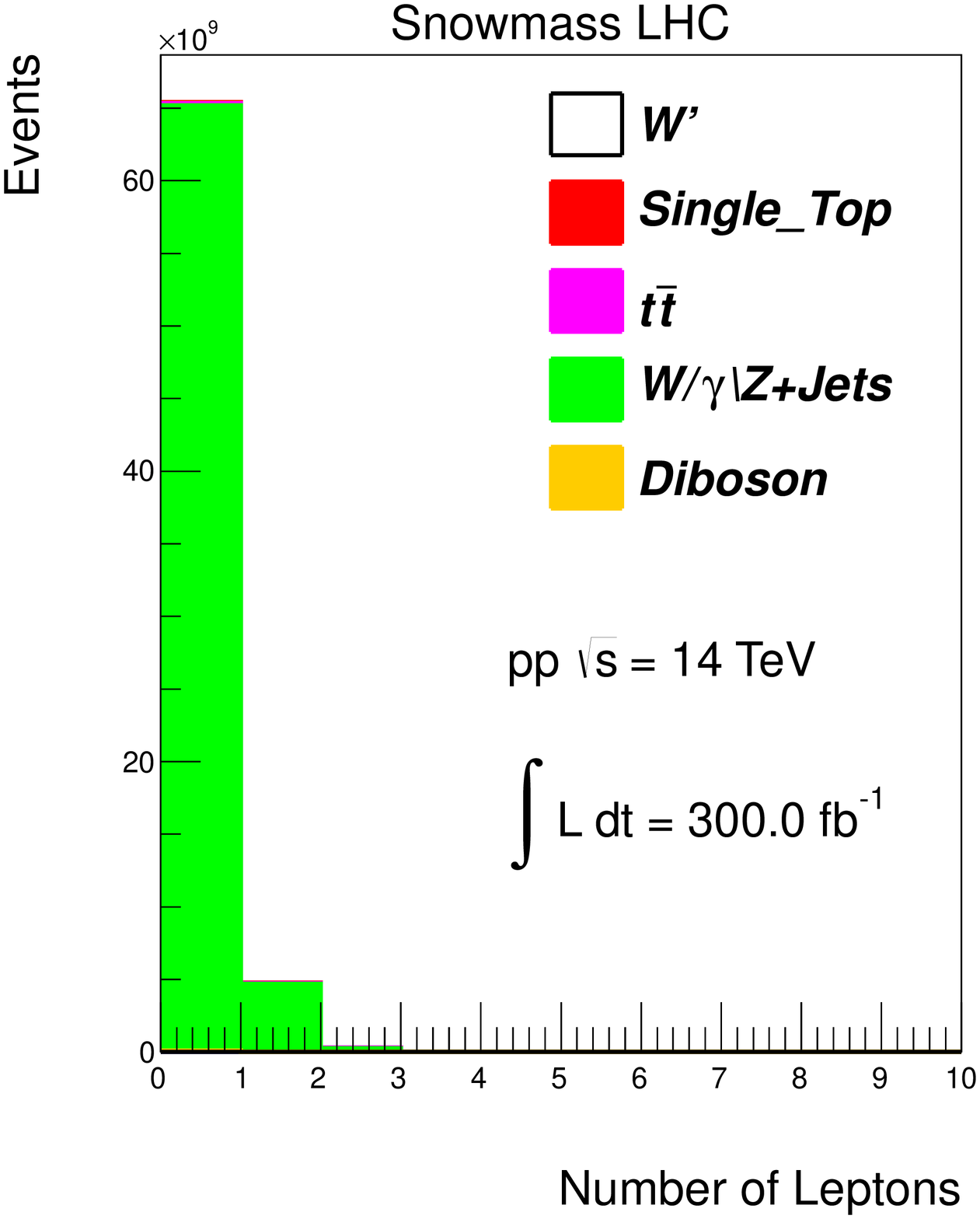}
    \label{fig:kin2a}
  }
  \subfigure[]{
    \includegraphics[width=0.37\textwidth]{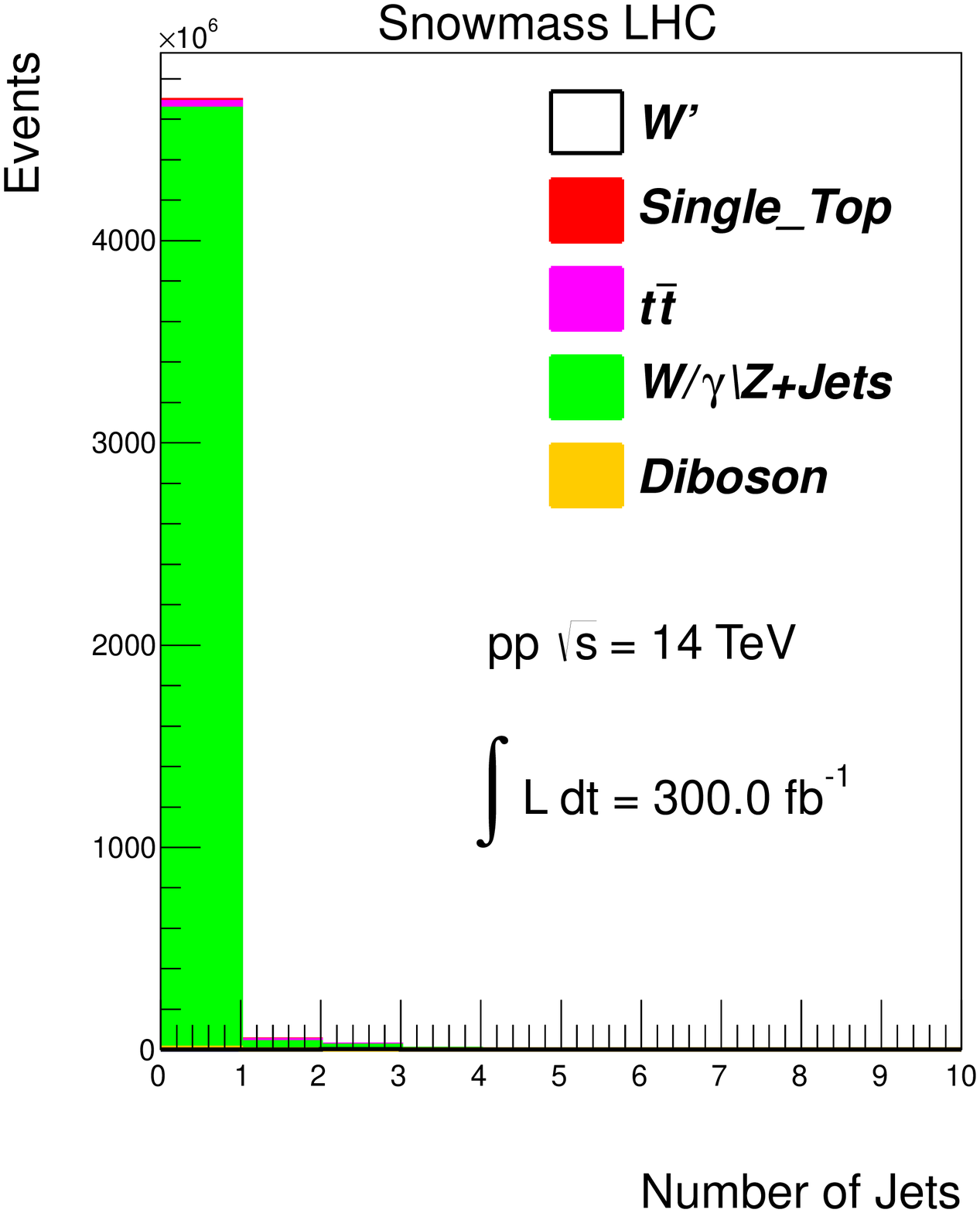}
    \label{fig:kin2b}
  }
  \subfigure[]{
    \includegraphics[width=0.37\textwidth]{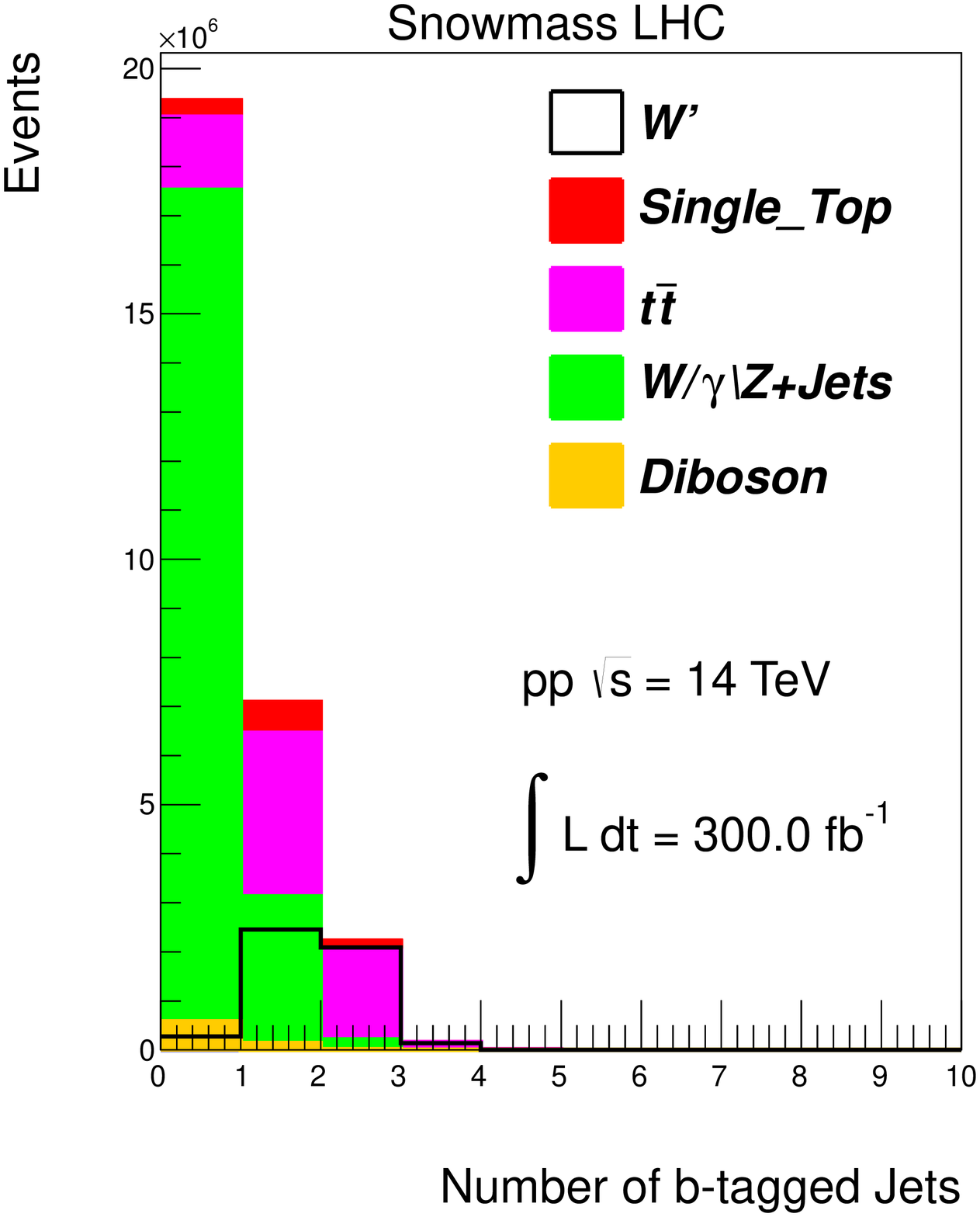}
    \label{fig:kin2c}
  } 
  \subfigure[]{
    \includegraphics[width=0.37\textwidth]{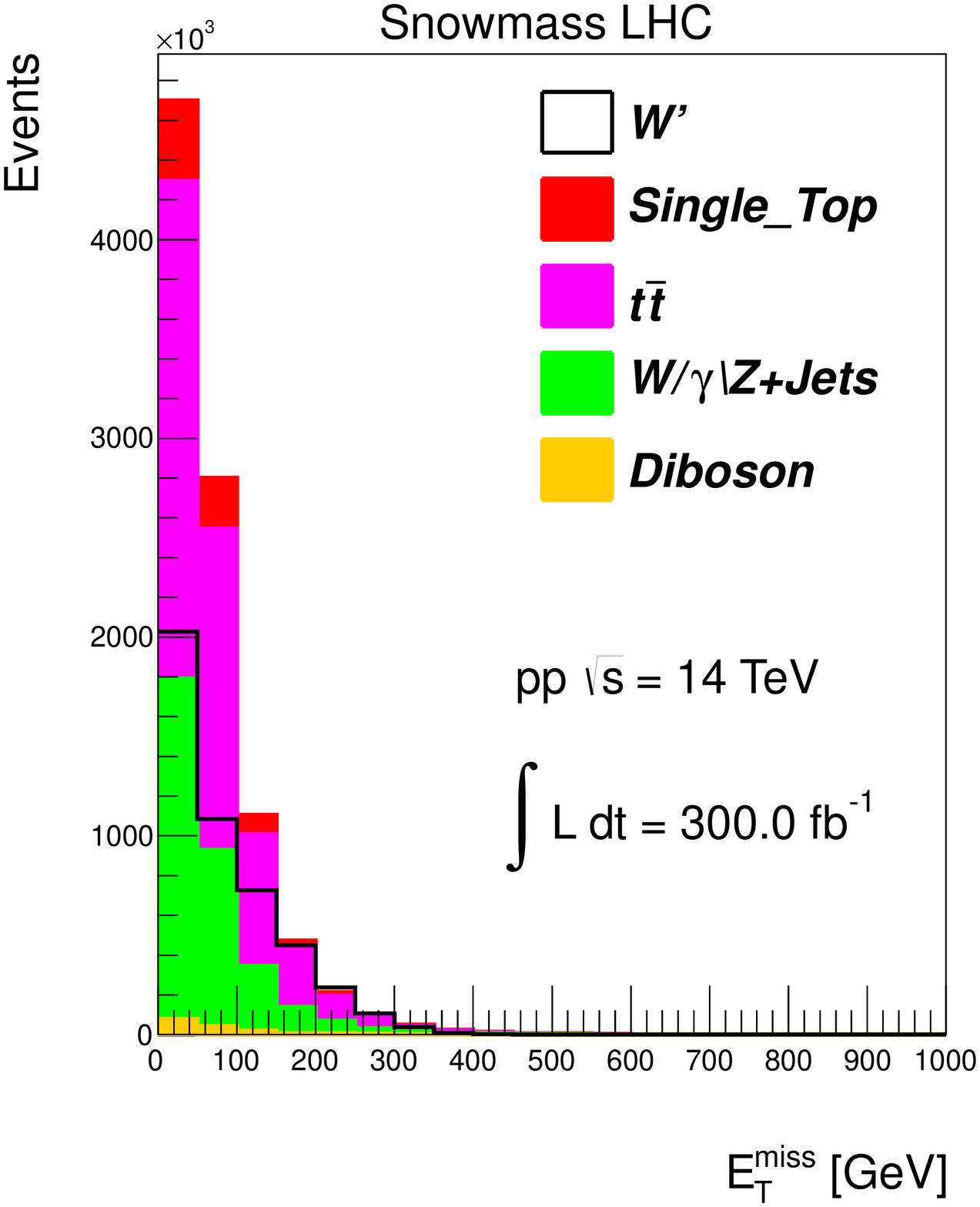}
    \label{fig:kin2d}
  } 
  \caption{Some kinematic distributions after all cuts leading up to the variable in question have 
    been applied: (a) exactly one lepton, (b) at least two jets, (c) at least one $b$-tagged jet, and (d) 
    $E_T^{miss}$ requirement.  The $W'$ contribution corresponds with a generated mass
    of 1~TeV for 300~$fb^{-1}$ at 14~TeV. This signal sample is scaled up by a factor of 10.}
  \label{fig:preselectionwpkin}
\end{figure}

\begin{figure}[H]
  \centering
  \subfigure[]{
    \includegraphics[width=0.37\textwidth]{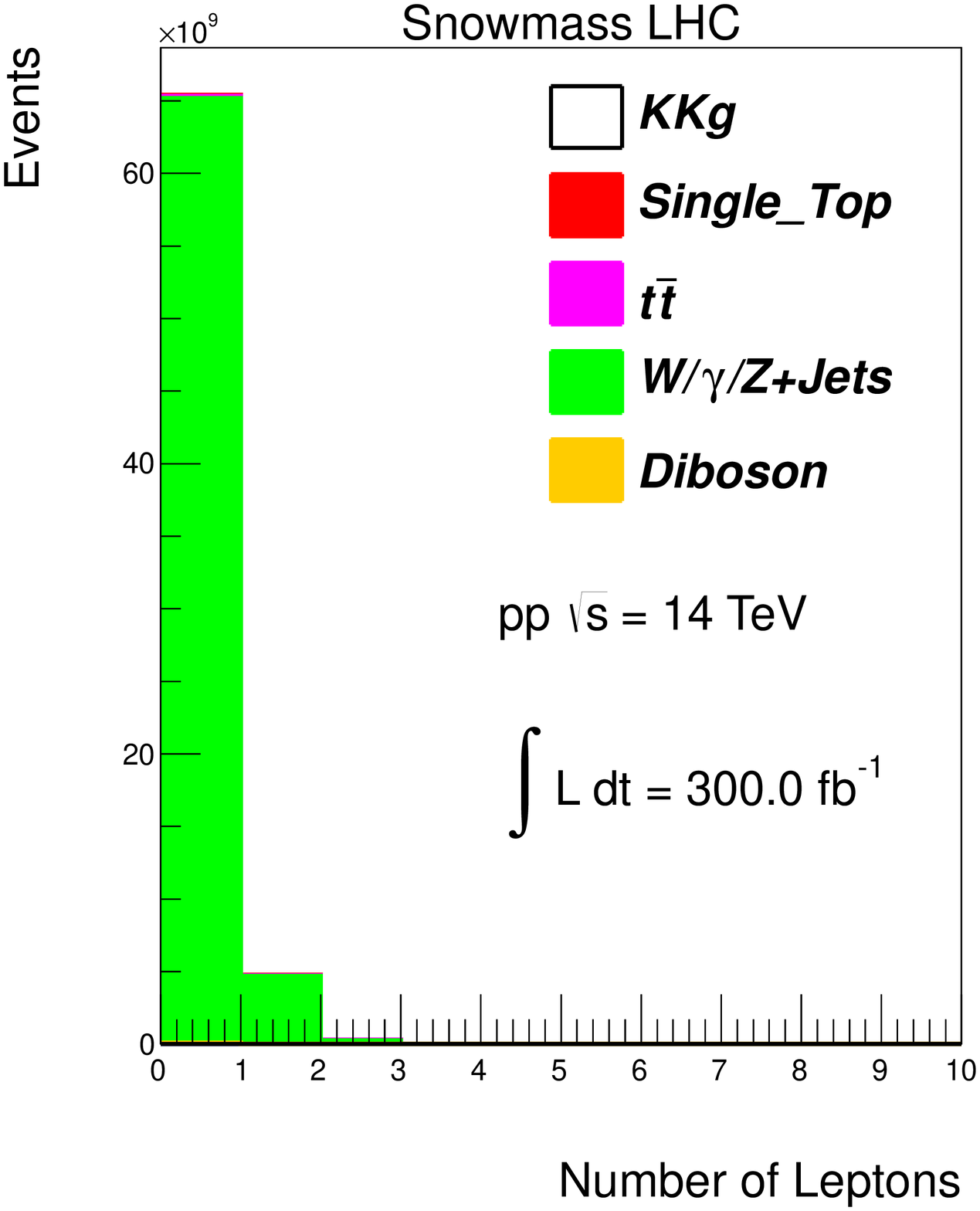}
    \label{fig:kin1a}
  }
  \subfigure[]{
    \includegraphics[width=0.37\textwidth]{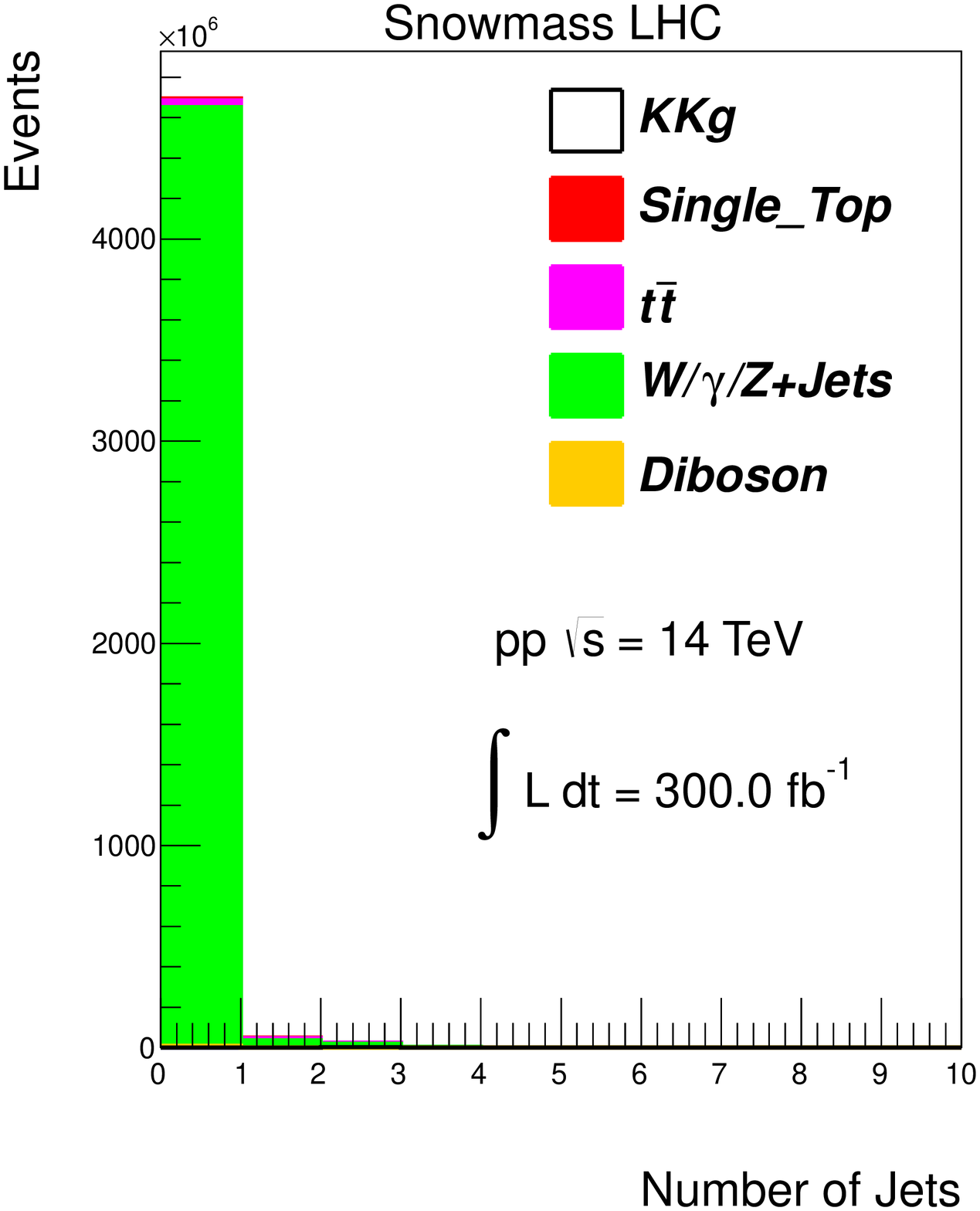}
    \label{fig:kin1b}
  }
  \subfigure[]{
    \includegraphics[width=0.37\textwidth]{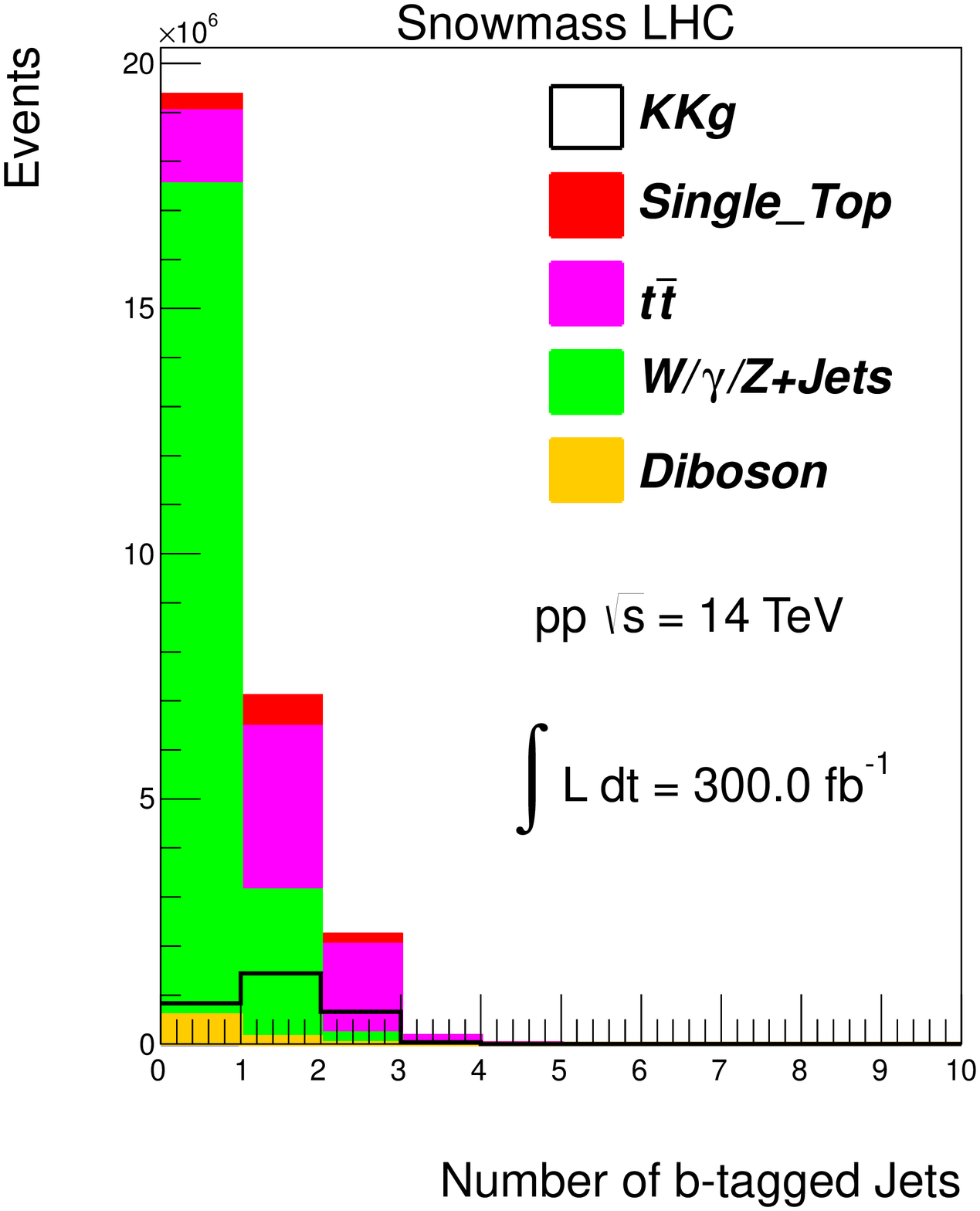}
    \label{fig:kin1c}
  } 
  \subfigure[]{
    \includegraphics[width=0.37\textwidth]{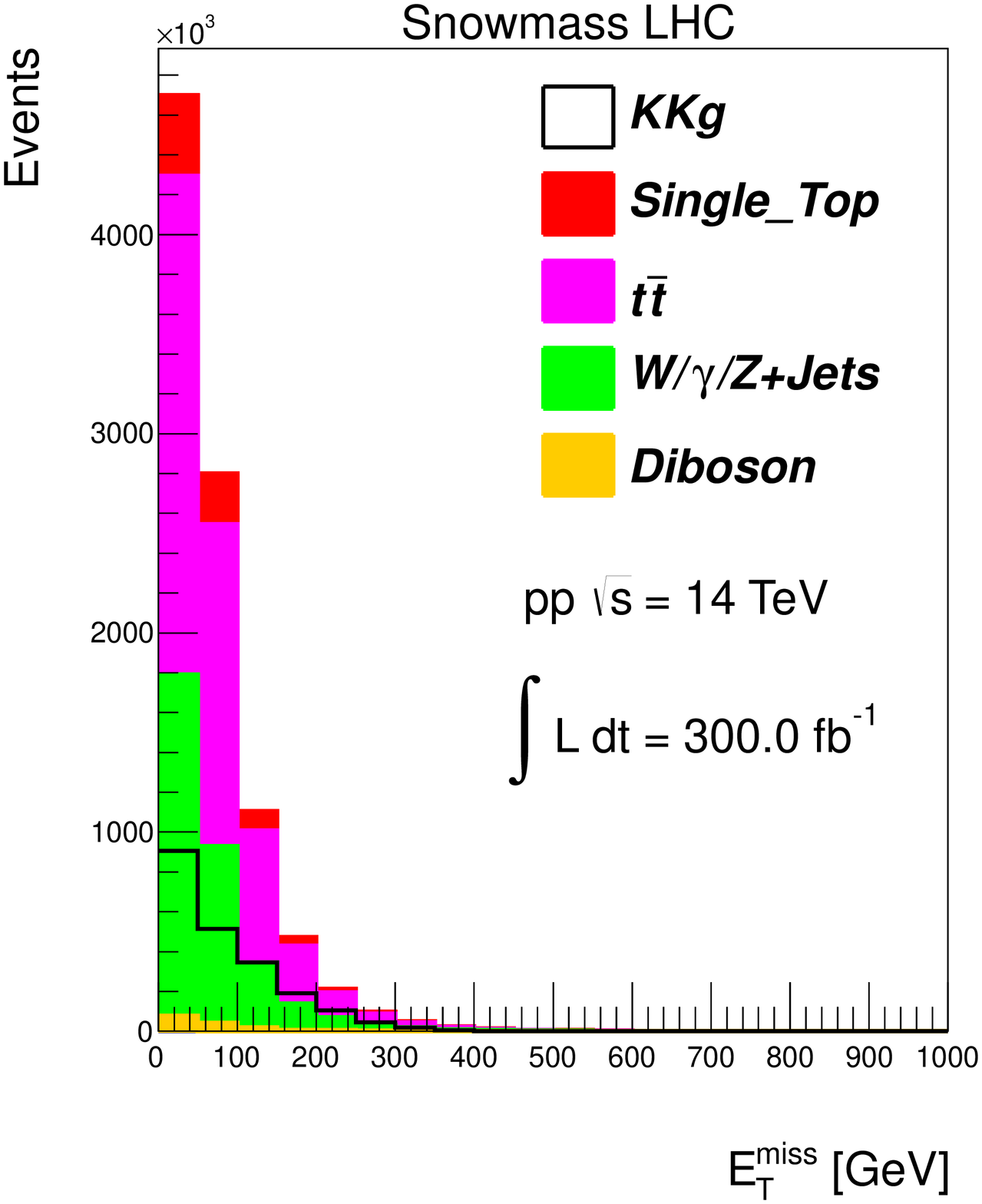}
    \label{fig:kin1d}
  } 
  \caption{Kinematic distributions after all cuts leading up to the variable in question have 
    been applied: (a) exactly one lepton, (b) at least two jets, (c) at least one $b$-tagged jet, and 
    (d) $E_T^{miss}$ requirement. The $KKg$ contribution corresponds with a generated mass
    of 1~TeV for 300~$fb^{-1}$ at 14~TeV. This signal sample is scaled up by a factor of 10.}
  \label{fig:preselectionkkgkin}
\end{figure}

Figures~\ref{fig:preselectionkkgkin} and~\ref{fig:preselectionwpkin} show kinematic
distributions for these preselection variables when cuts are applied in the order given in
Eq.~\ref{eq:basiccut}, for $KKg$ and $W'$, respectively. 
~

\section{Event selection}
\label{sec:selection}

To improve the signal to background ratio, events are required to pass the following additional
selection cuts in the $KKg$ analysis:
\begin{center}
\begin{eqnarray}
\textrm{Lepton with } p_{T}^{\ell} \geq 100\,{\rm GeV}, \nonumber \\
\textrm{Jet with } p_{T}^{j} \geq 300\,{\rm GeV}, \nonumber \\
\textrm{Number of Subjets } \leq 2, \nonumber \\
\textrm{Exactly 1 B-Tagged Jet}, \nonumber \\
\textrm{Exactly 1 Non-B-Tagged Jet}, \nonumber \\
\textrm{Top Mass} \leq 200\,{\rm GeV}, \nonumber \\
\textrm{Reconstructed $KKg$ Mass} \geq \textrm{Generated Value.} \nonumber \\
\label{eq:extracutsKKg}
\end{eqnarray}
\end{center}

The cuts are chosen from a larger set of available variables and cuts through a systematic
approach. For each cut on a given variable, the integral of the signal over the square root
of the background is computed. In addition, since the event yields in this analysis are large,
the systematic uncertainty on the background (10\%) is taken into account by
defining the figure of merit ($F.O.M$) as
\begin{eqnarray}
F.O.M = \frac{S}{\sqrt{B + (0.1 \cdot B)^2}}\, ,
\end{eqnarray}
where $S$ ($B$) is the number of expected signal (background) events passing the cut.

\begin{figure}[h]
  \centering
  \subfigure[]{
    \includegraphics[width=0.31\textwidth]{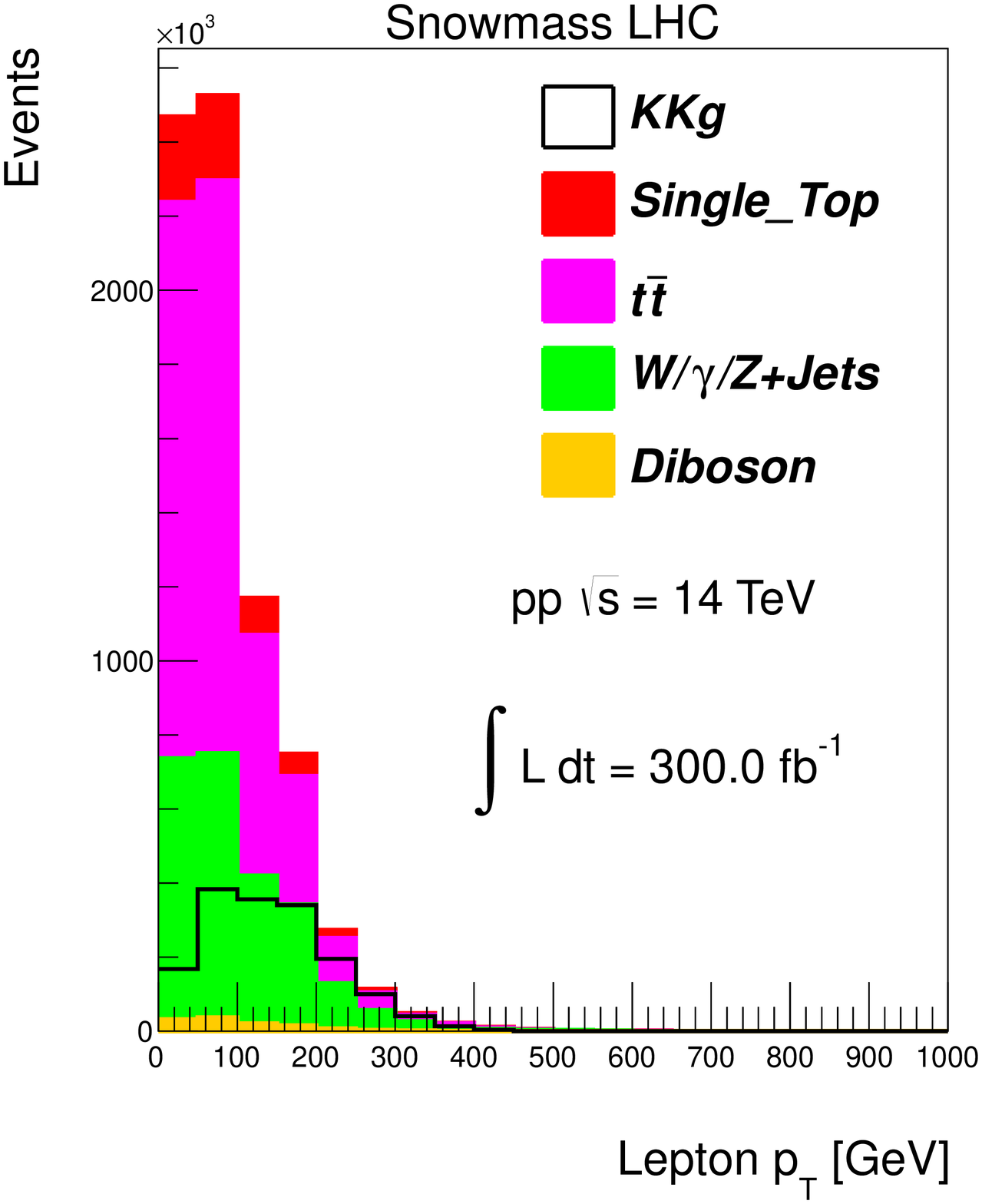}
    \label{fig:kin3a}
  }
  \subfigure[]{
    \includegraphics[width=0.31\textwidth]{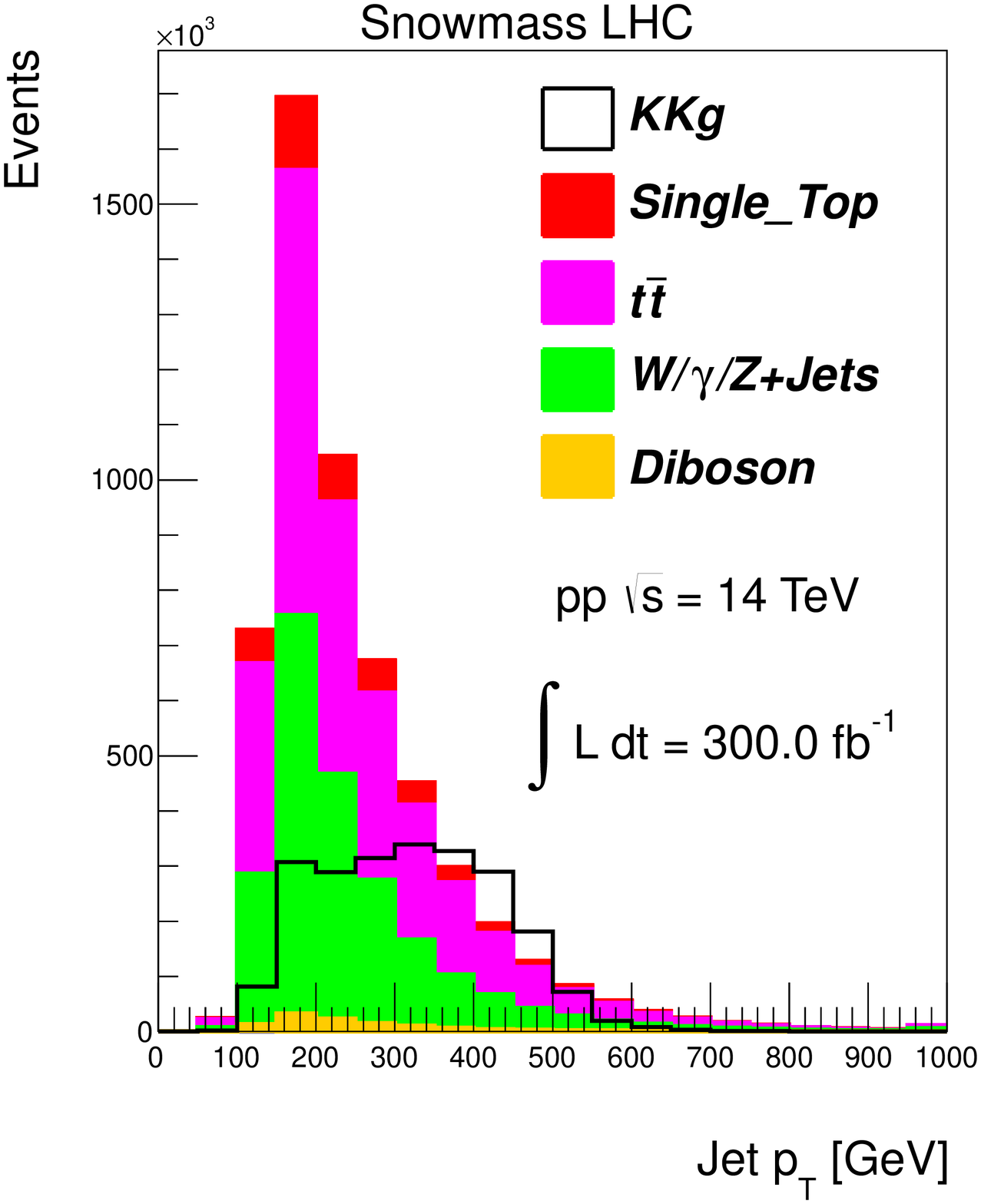}
    \label{fig:kin3b}
  }
  \subfigure[]{
    \includegraphics[width=0.31\textwidth]{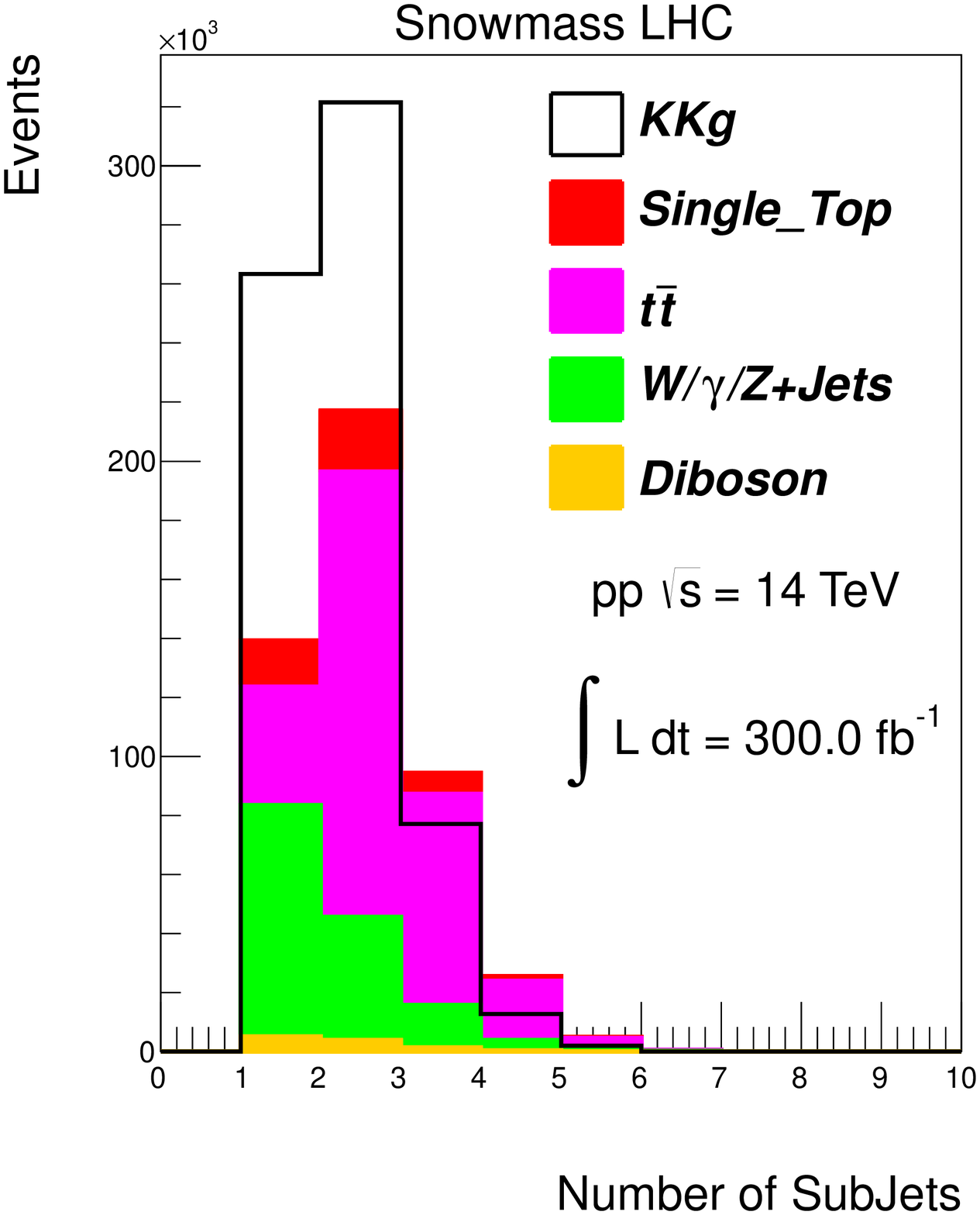}
    \label{fig:kin3c}
  } 
  \subfigure[]{
    \includegraphics[width=0.31\textwidth]{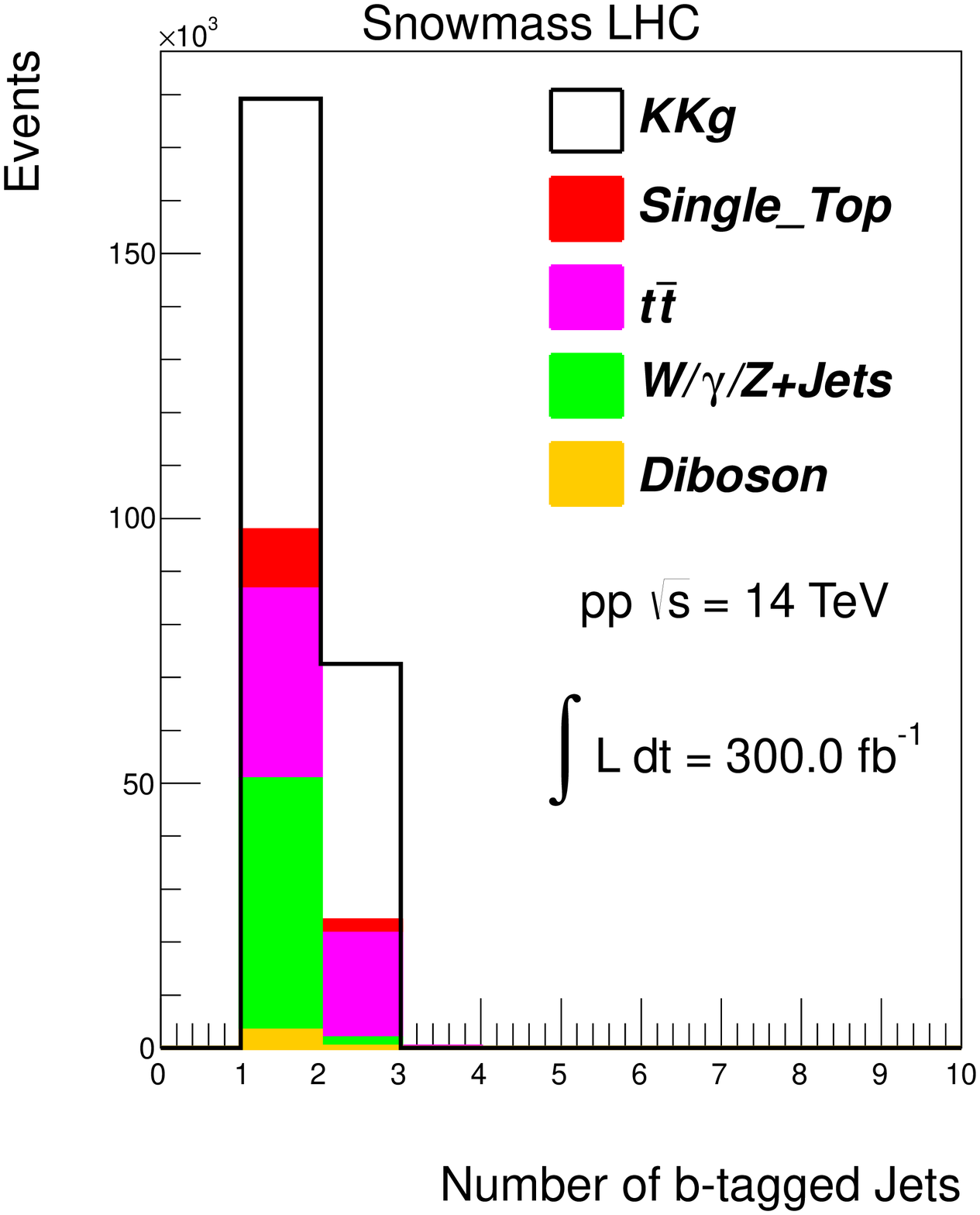}
    \label{fig:kin3d}
  } 
 \subfigure[]{
    \includegraphics[width=0.31\textwidth]{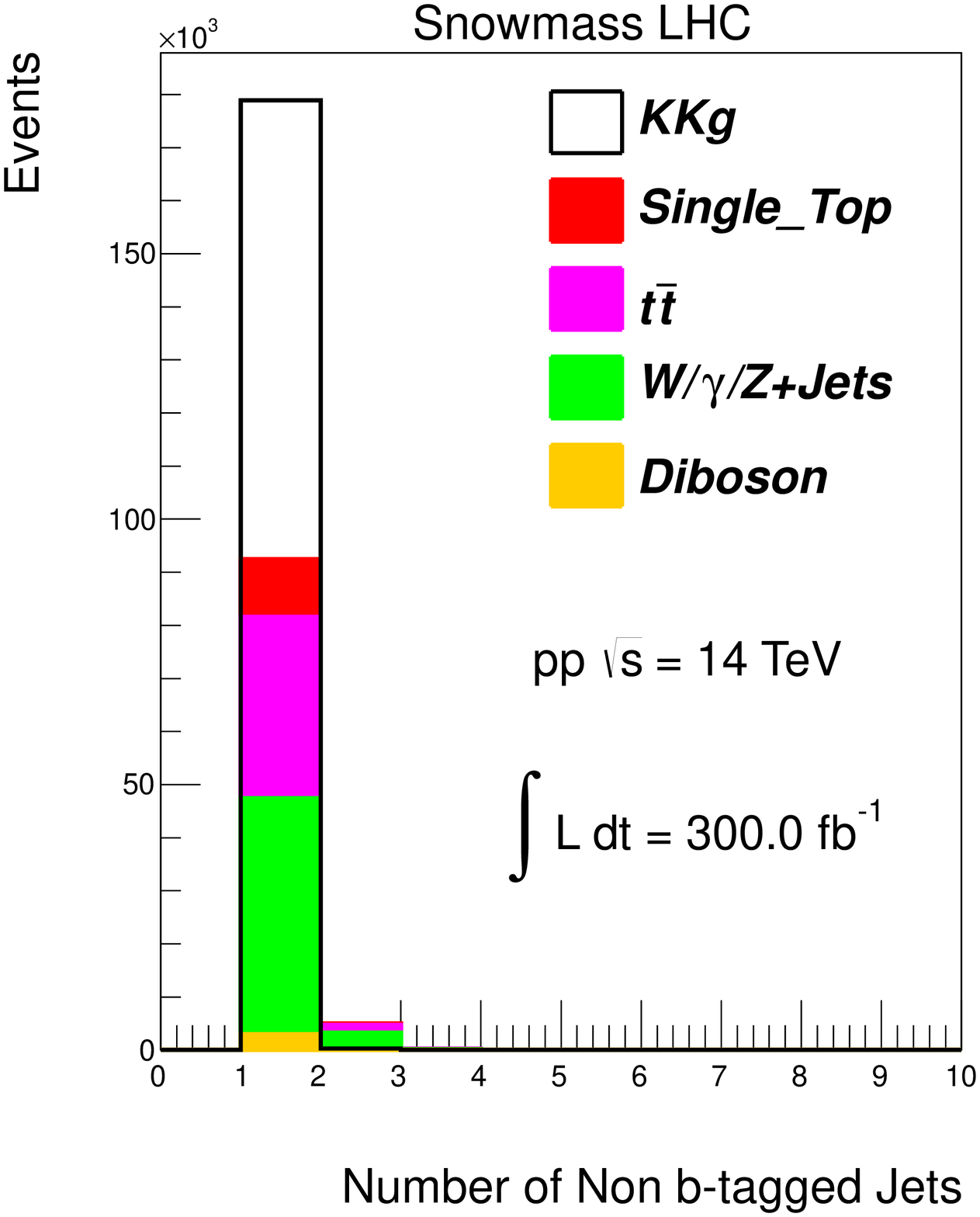}
    \label{fig:kin3e}
  } 
  \subfigure[]{
    \includegraphics[width=0.31\textwidth]{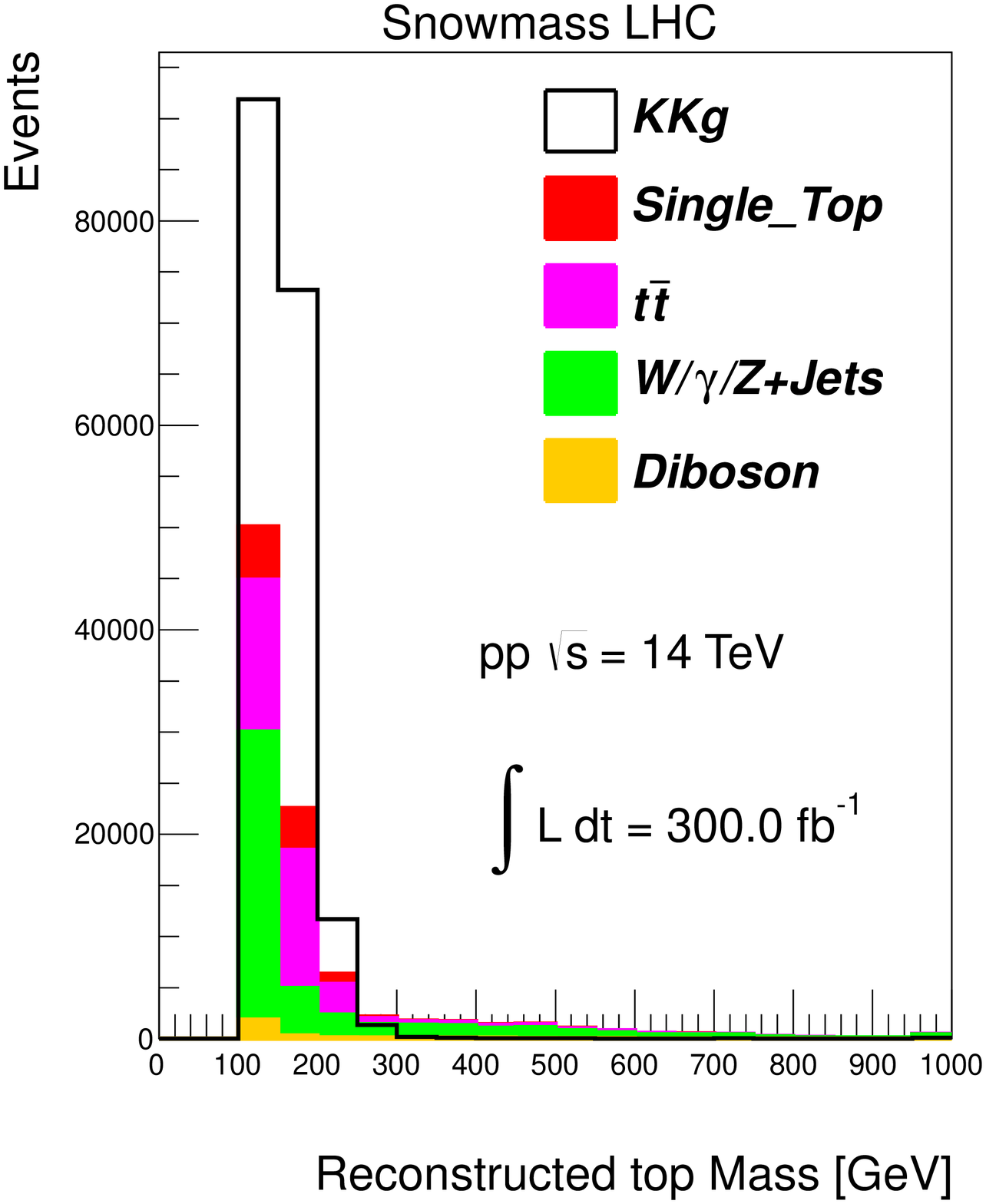}
    \label{fig:kin3f}
  } 
  \vspace{-.2cm}
  \caption{Kinematic distributions after all cuts leading up to the variable in question have 
    been applied: (a) lepton $p_T$, (b) jet $p_T$, (c) number of subjets $\leq$ 2, (d) exactly one $b$-tagged 
    jet, (e) exactly one non-$b$-tagged jet, and (f) top mass. The $KKg$ contribution corresponds with a generated mass
    of 1~TeV for 300~$fb^{-1}$ at 14~TeV. This signal sample is scaled up by a factor of 10.}
 \label{fig:kinematicKKG20}
\end{figure}

Figure~\ref{fig:kinematicKKG20} shows the kinematic distributions for these $KKg$ analysis
variables for events that pass the preselection cuts (Eq.~\ref{eq:basiccut}) and the cuts from
Eq.~\ref{eq:extracutsKKg} up to the variable in question. 

\begin{figure}[h]
  \centering
  \subfigure[]{
    \includegraphics[width=0.33\textwidth]{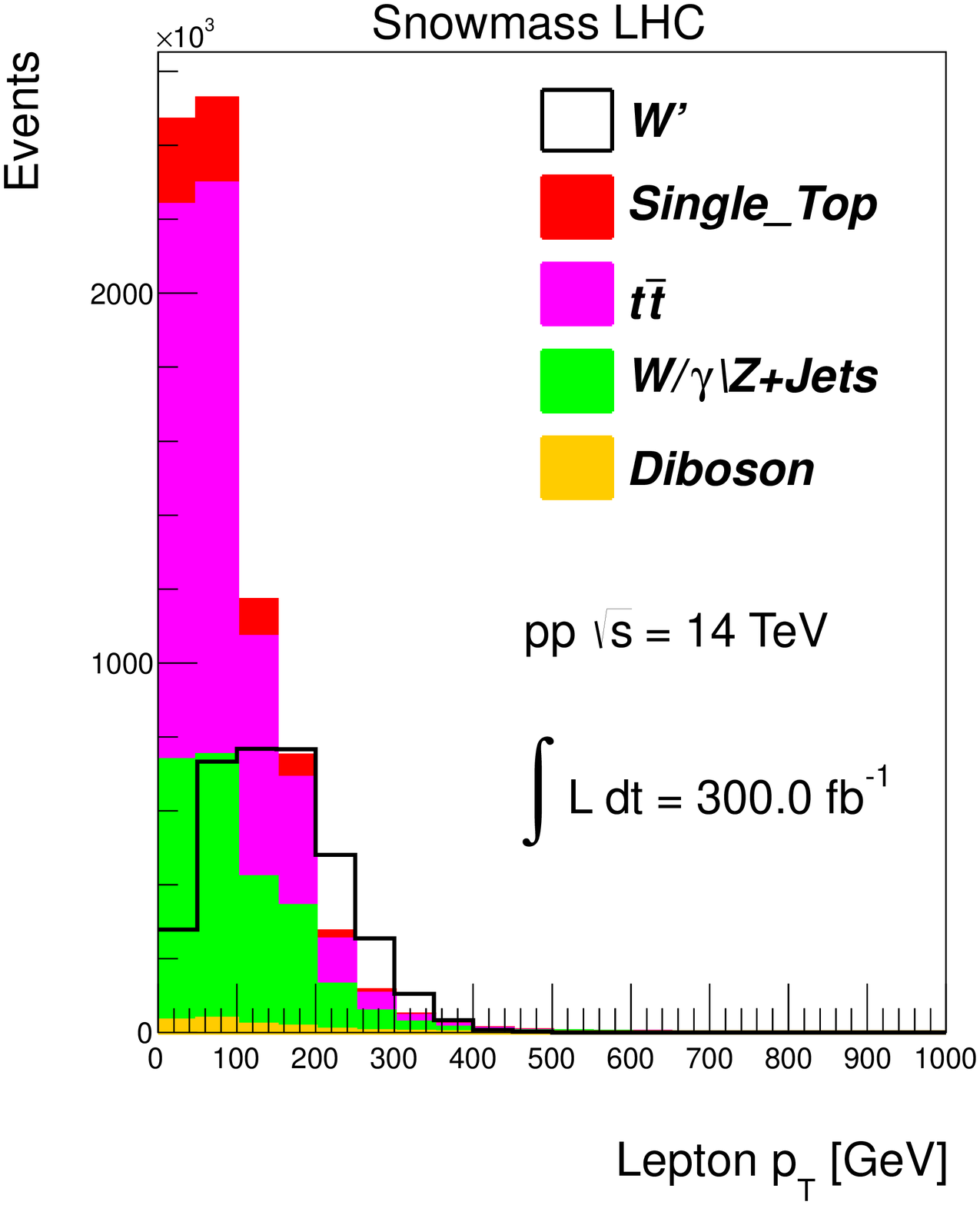}
    \label{fig:kin4a}
  }
  \subfigure[]{
    \includegraphics[width=0.33\textwidth]{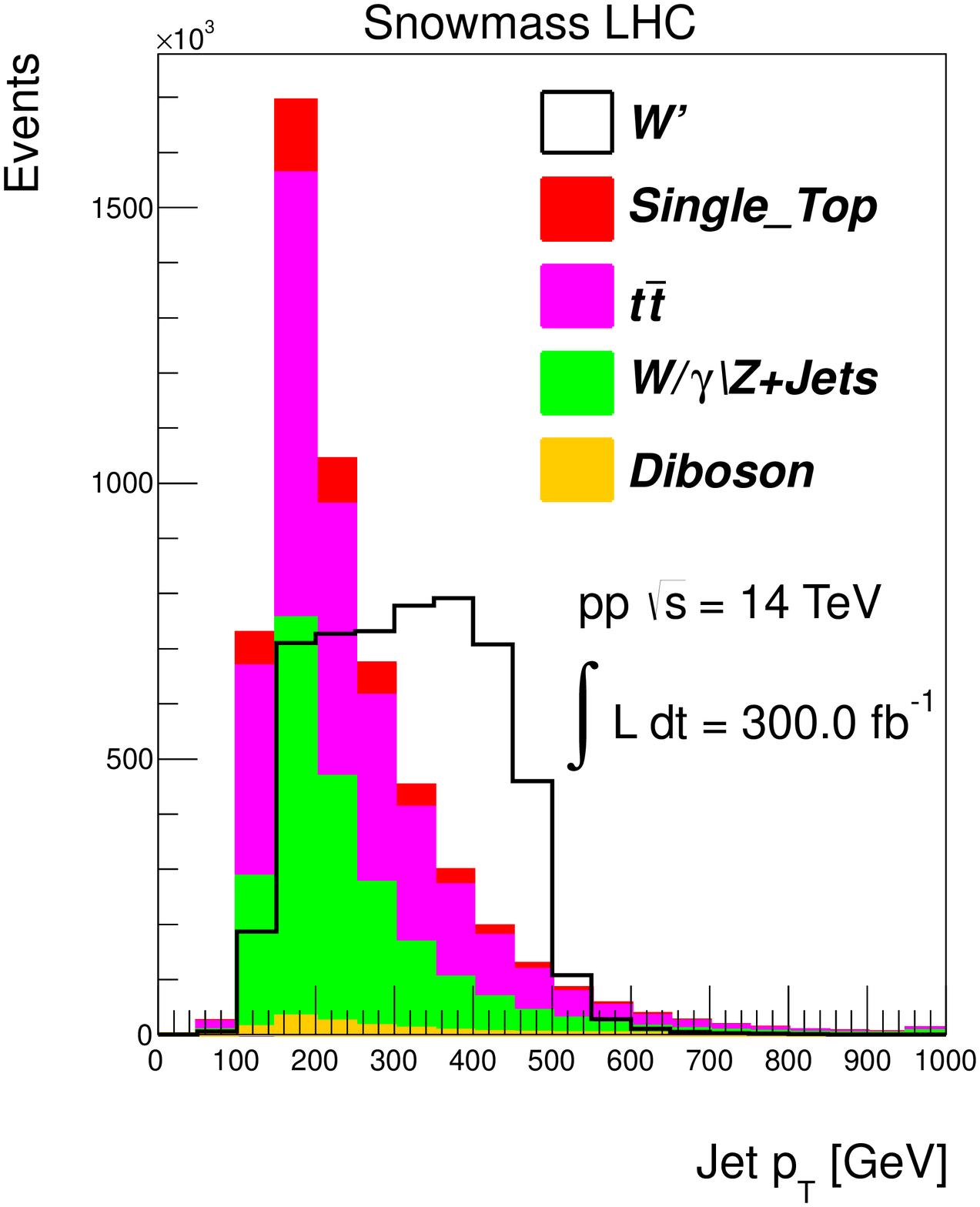}
    \label{fig:kin4b}
  }
  \subfigure[]{
    \includegraphics[width=0.33\textwidth]{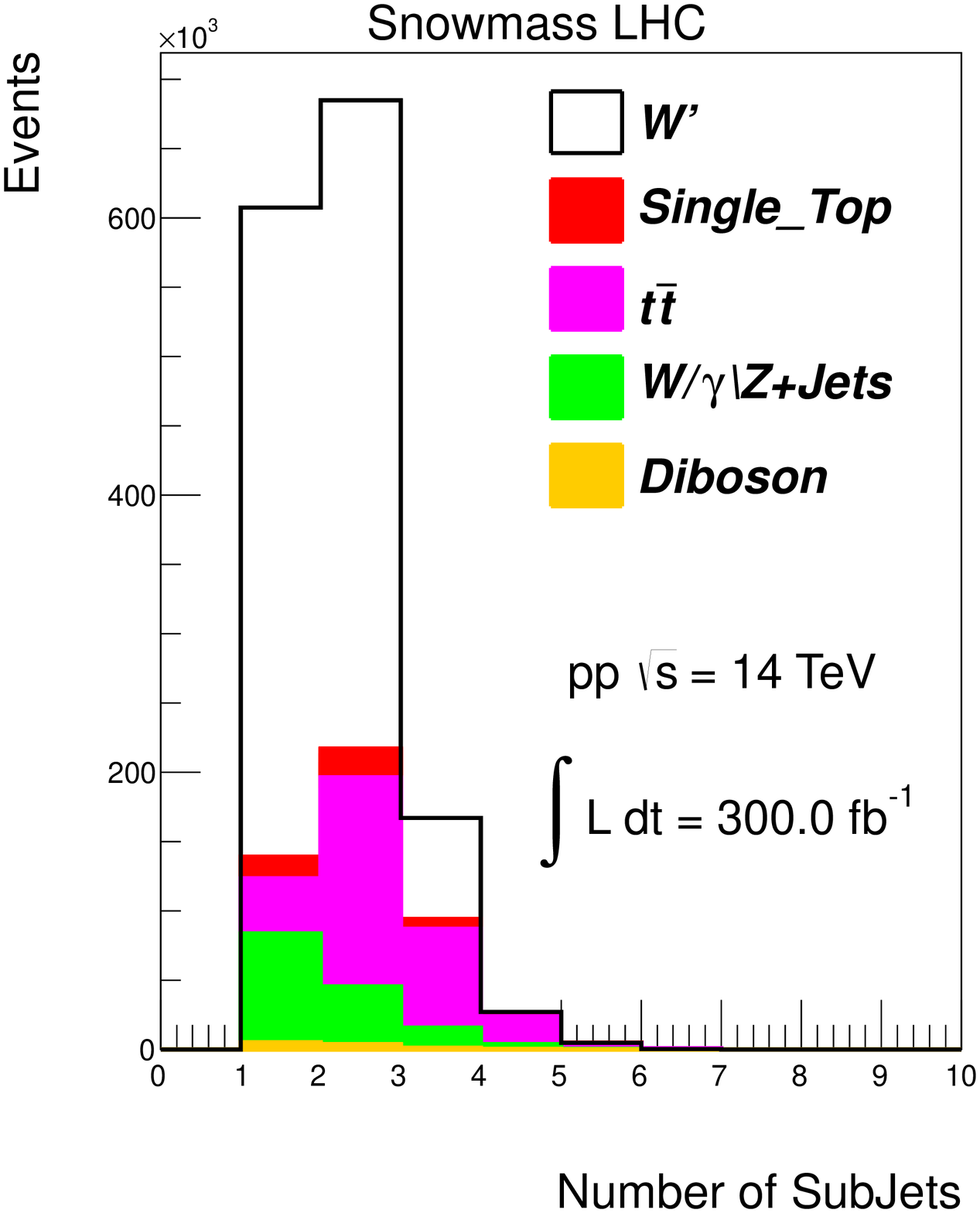}
    \label{fig:kin4c}
  } 
  \subfigure[]{
    \includegraphics[width=0.33\textwidth]{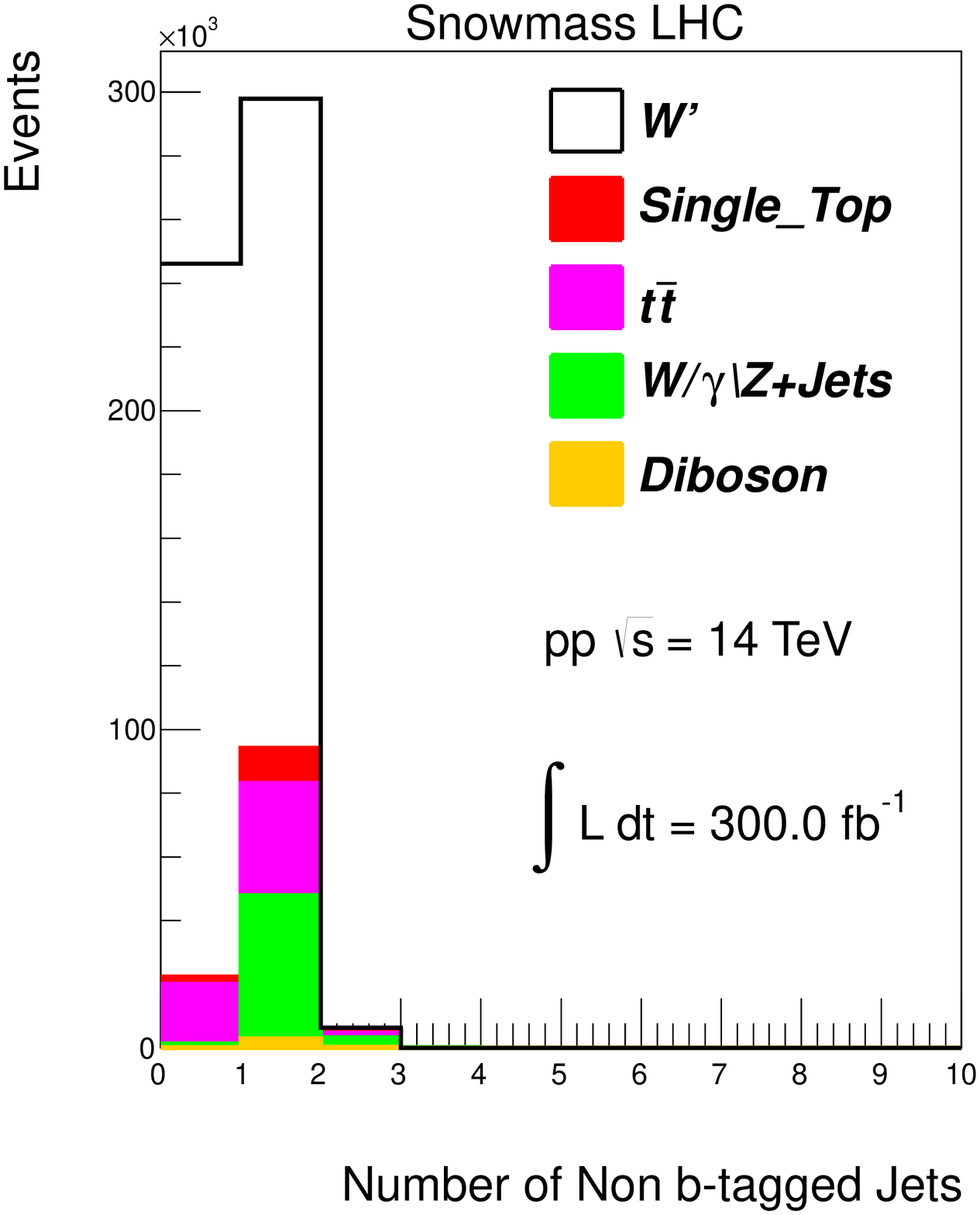}
    \label{fig:kin4d}
  } 
  \caption{Some kinematic distributions after all cuts leading up to the variable in question have 
    been applied: (a) stricter lepton $p_T$ cut, (b) stricter jet $p_T$ cut, (c) number of subjets $\leq$ 2, 
    and (d) no non-$b$-tagged jets. The $W'$ contribution corresponds with a generated mass
    of 1~TeV for 300~$fb^{-1}$ at 14~TeV. This signal sample is scaled up by a factor of 10.}
  \label{fig:kinematicWP}
\end{figure}

Similar, though not identical, cuts are used for the $W'$ analysis:
\begin{center}
  \begin{eqnarray}
    \textrm{Lepton with } p_{T}^{\ell} \geq 100\,{\rm GeV}, \nonumber \\
    \textrm{All Jets with } p_T^j \geq 300\,{\rm GeV}, \nonumber \\
    \textrm{Number of Subjets } \leq 2, \nonumber \\
    \textrm{Exactly 0 Non-B-Tagged Jets}, \nonumber \\
    \textrm{Reconstructed $W'$ Mass } \geq \textrm{Generated Value.} \nonumber \\
    \label{eq:extracutsW}
  \end{eqnarray}
\end{center}

Figure~\ref{fig:kinematicWP} shows the kinematic distributions for these $W'$ analysis
variables for events that pass the preselection cuts (Eq.~\ref{eq:basiccut}) and the cuts from
Eq.~\ref{eq:extracutsW} up to the variable in question. 
~

In each case, the neutrino is reconstructed from the $E_T^{miss}$, $E_T^{mis} ~\phi$, lepton $p_T$, 
lepton $\eta$, and lepton $\phi$ using a $W$~boson mass constraint, which results in a 
quadratic equation that yields either one solution, two solutions, or imaginary solutions for the neutrino 
$p_Z$. In the case of two solutions, the smaller $|p_Z|$ solution of the two is chosen. In the case of imaginary solutions, 
the neutrino $p_T$ is scaled such that a solution is obtained with the $p_Z = 0$.
The $W$ is reconstructed by adding the reconstructed neutrino to the lepton. The
top is reconstructed using the reconstructed $W$ and the jet that gives a top mass closest to 172.0~GeV. 
The $KKg$ and $W'$ are both reconstructed using the reconstructed $W$ and the two leading jets.

\section{Results}
\label{sec:results}

\begin{figure}[H]
  \begin{center}
    \subfigure[]{
      \includegraphics[width=0.35\textwidth]{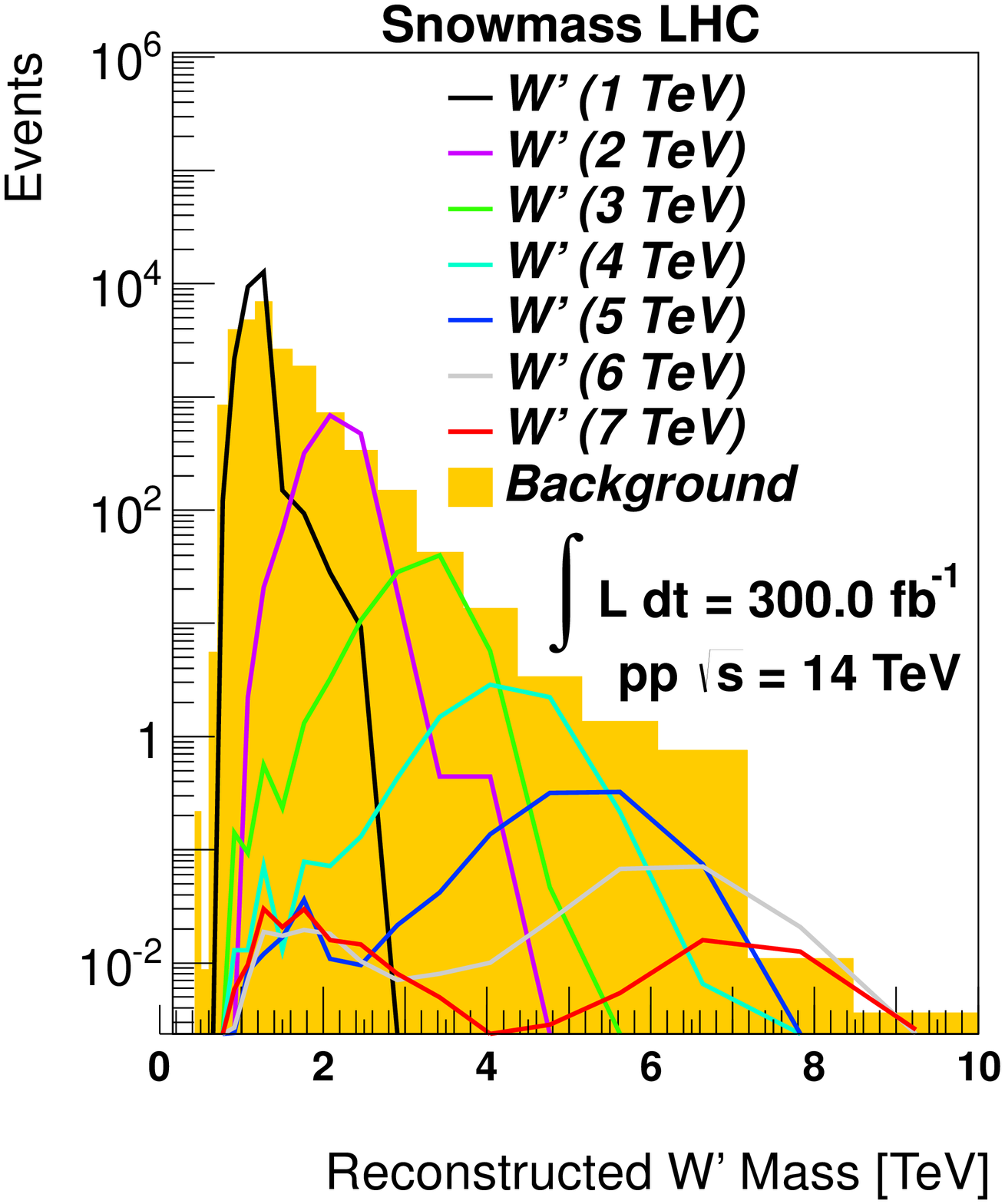}
      \label{fig:wma}
    }
    \subfigure[]{
      \includegraphics[width=0.35\textwidth]{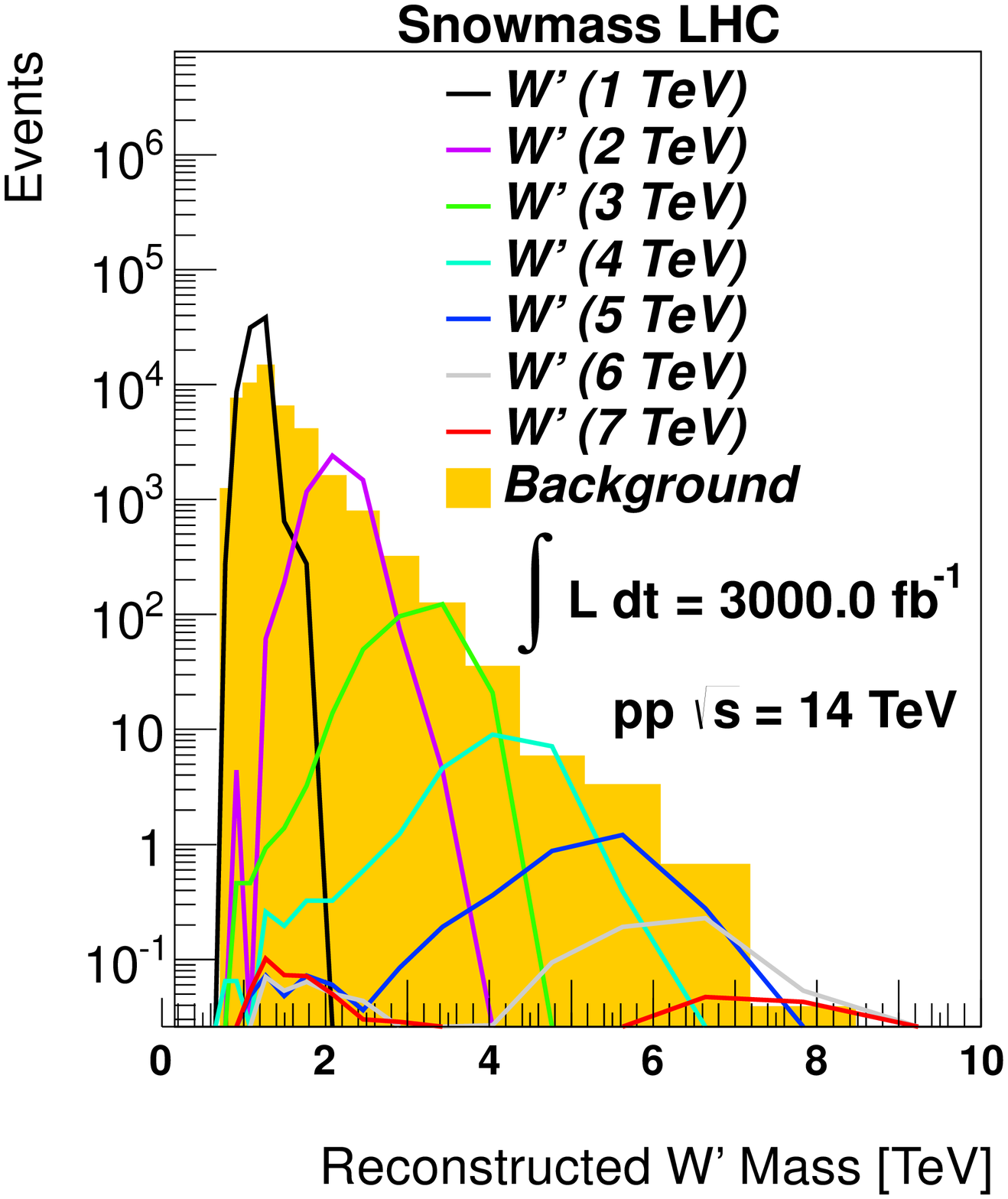}
      \label{fig:wmb}
    }
    \subfigure[]{
      \includegraphics[width=0.35\textwidth]{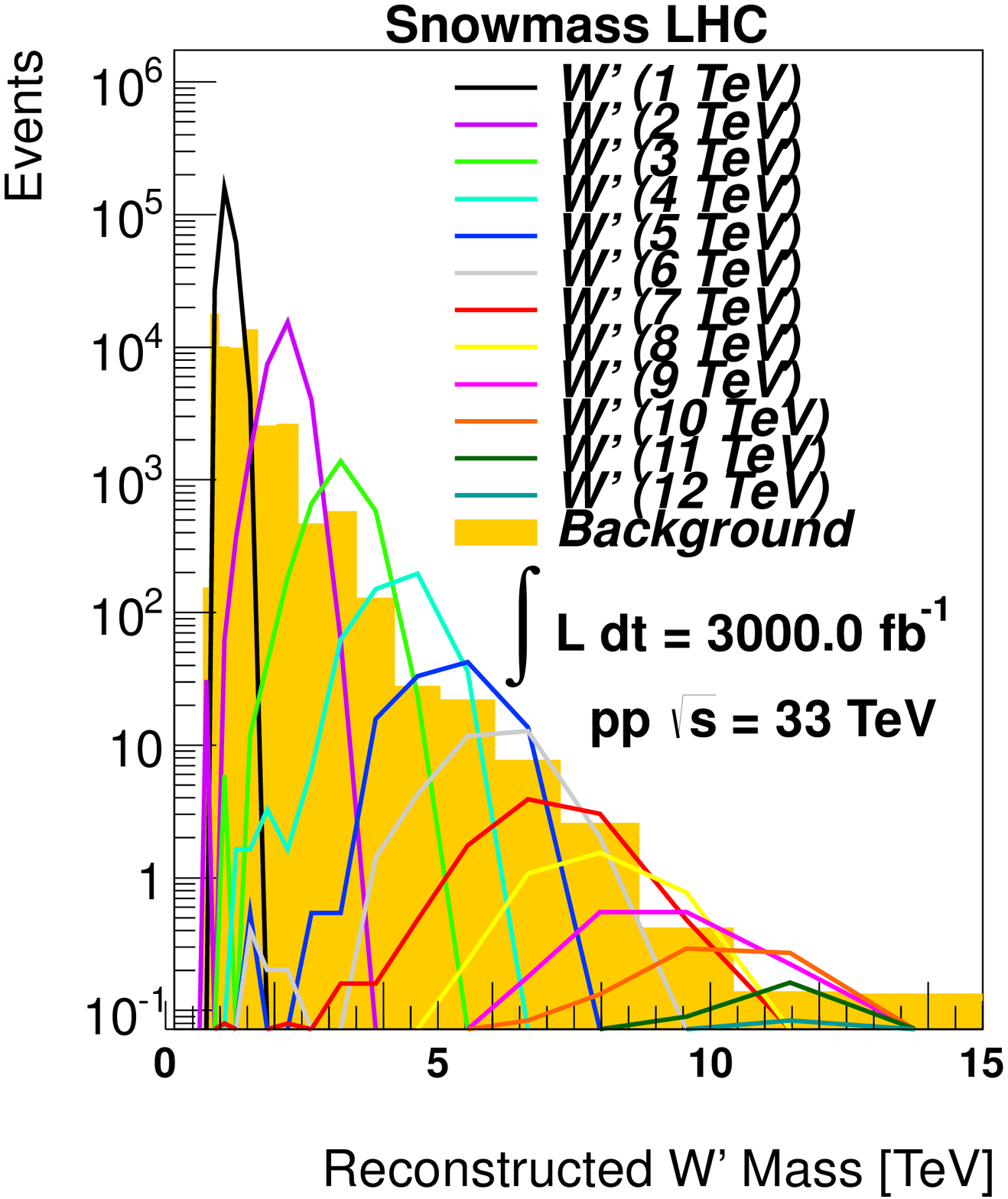}
      \label{fig:wmc}
    }
    \caption{The reconstructed $W'$ masses: (a) at 14~TeV and 300~$fb^{-1}$ luminosity,
      (b) at 14~TeV and 3000~$fb^{-1}$ luminosity , (c) at 33~TeV and 3000~$fb^{-1}$ luminosity.}
    \label{fig:kinematicswp}
  \end{center}
\end{figure}

\begin{figure}[H]
  \begin{center}
    \subfigure[]{
      \includegraphics[width=0.35\textwidth]{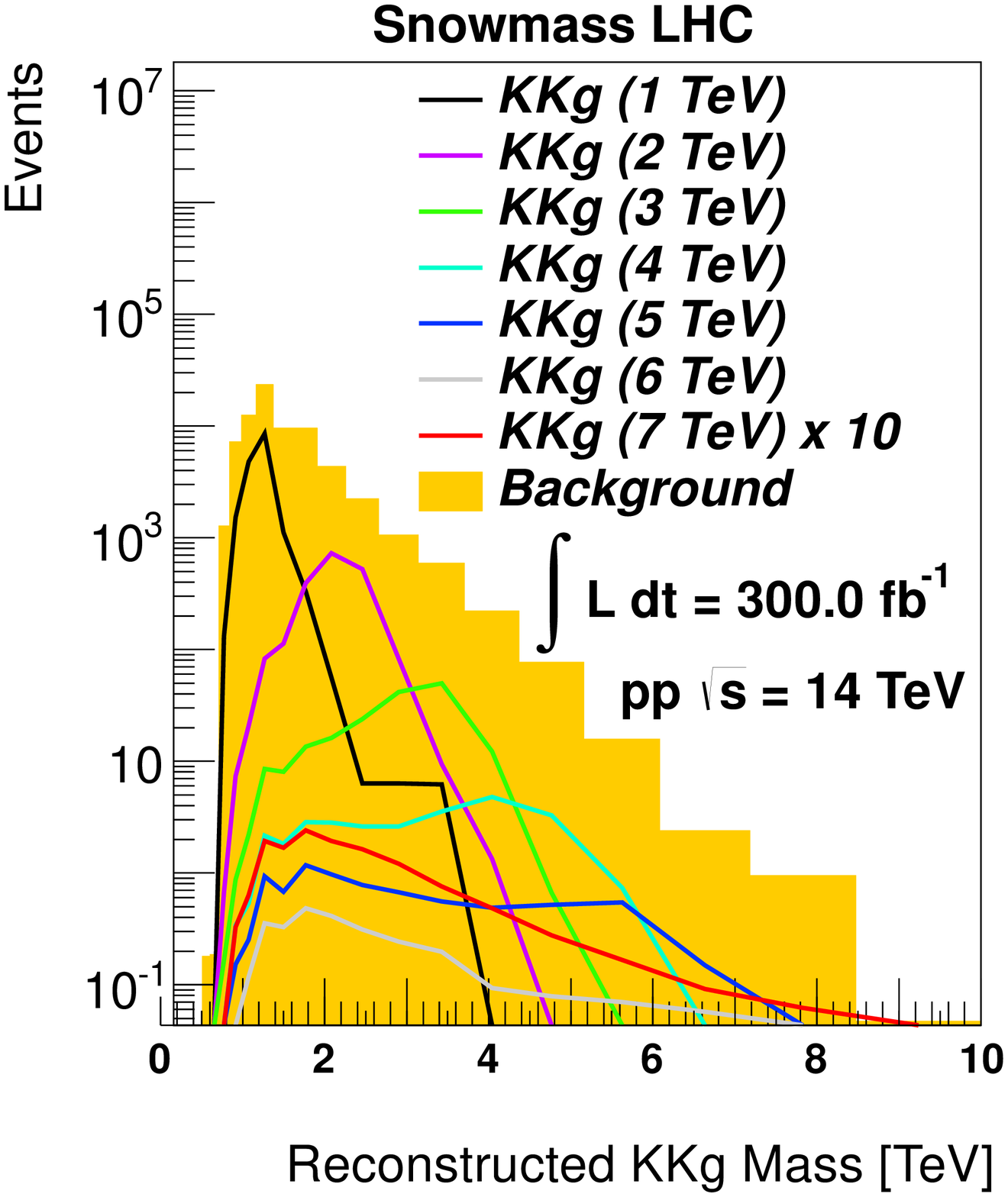}
      \label{fig:kkgma}
   }
    \subfigure[]{
      \includegraphics[width=0.35\textwidth]{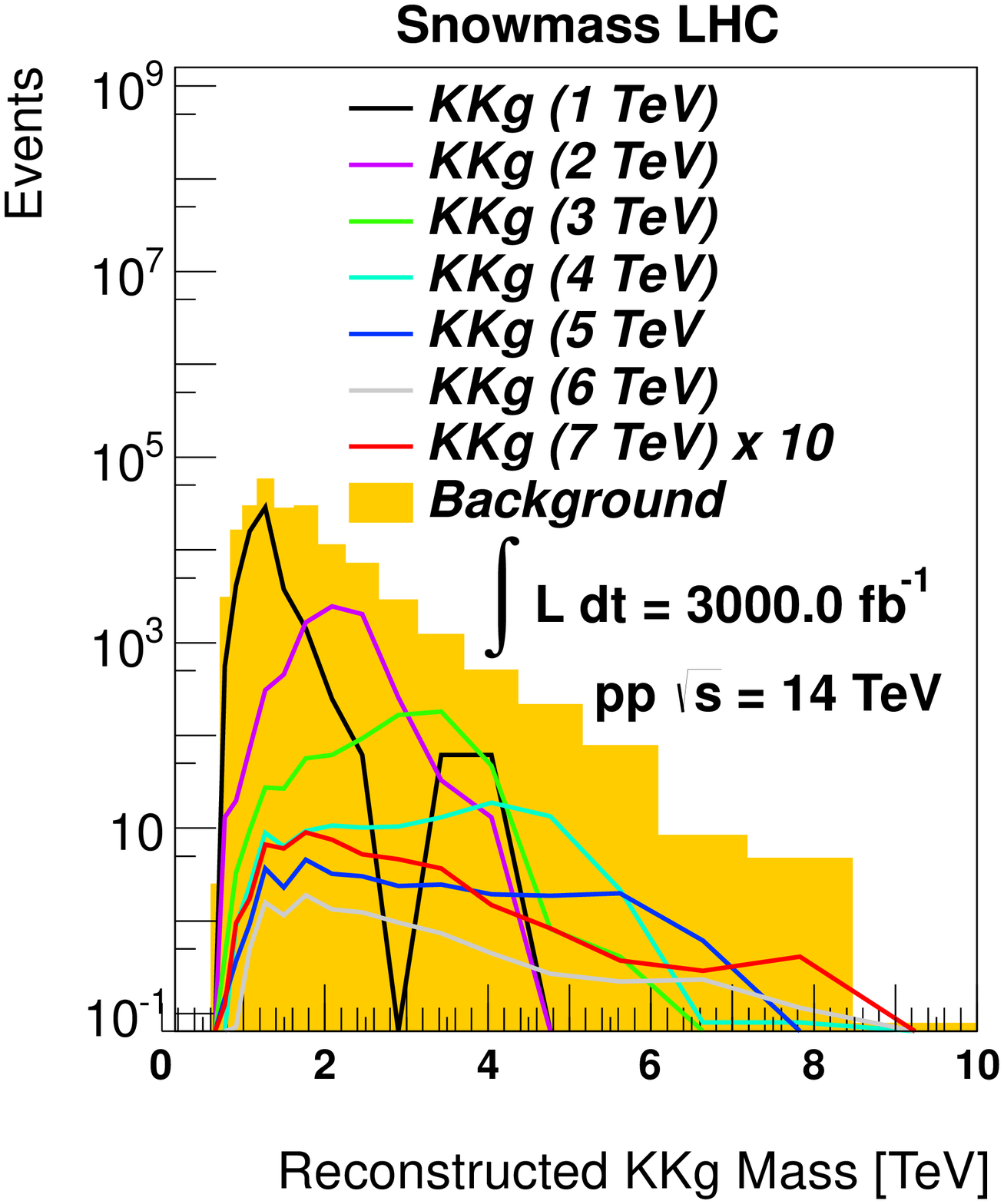}
      \label{fig:kkgmb}
    }
    \subfigure[]{
      \includegraphics[width=0.35\textwidth]{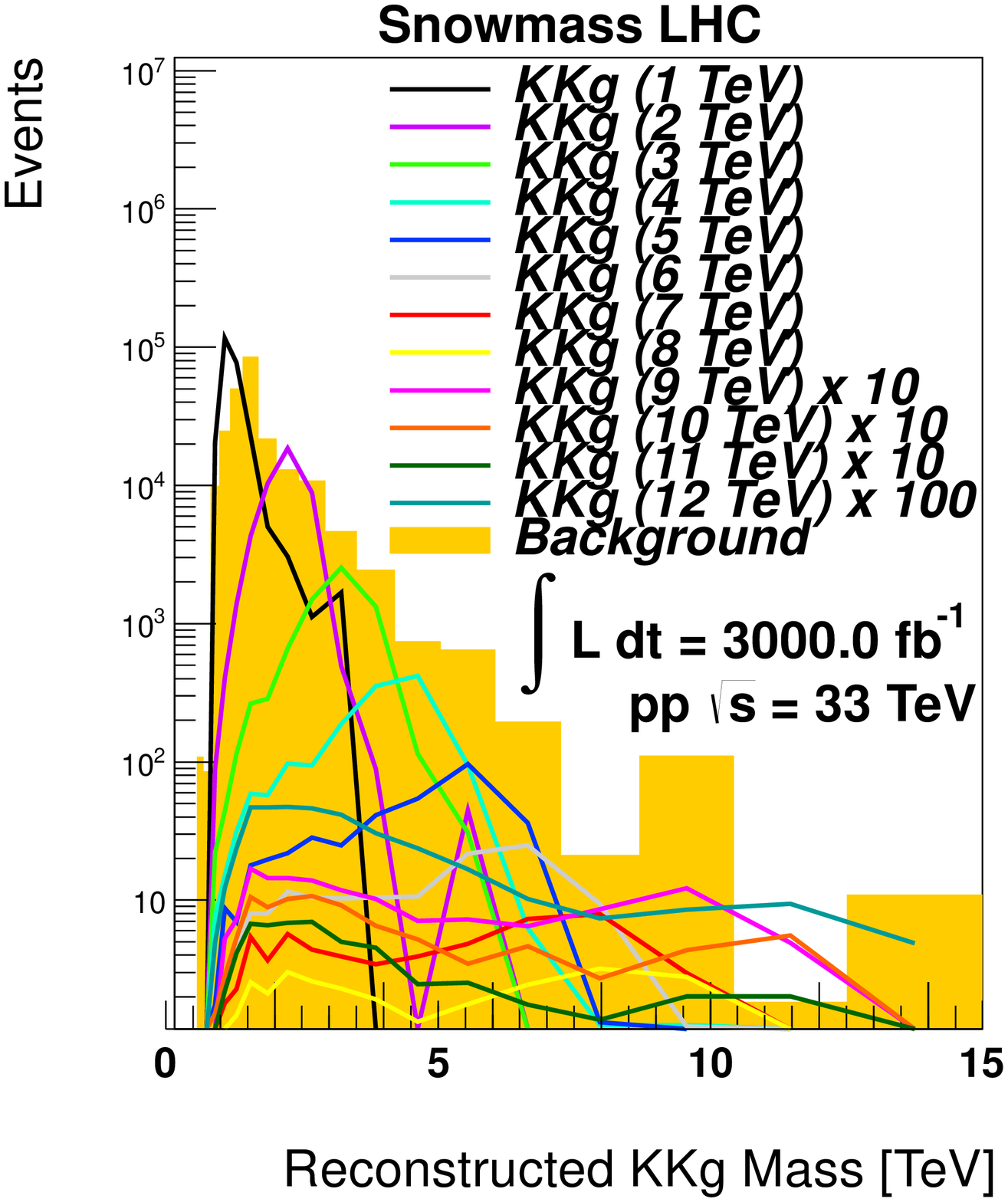}
      \label{fig:kkgmc}
    }
    \caption{The reconstructed KKg mass: (a) at 14~TeV and 300~$fb^{-1}$ luminosity,
      (b) at 14~TeV and 3000~$fb^{-1}$ luminosity, (c) at 33~TeV and 3000~$fb^{-1}$ luminosity.}
    \label{fig:kinematicskkg20}
  \end{center}
\end{figure}

The reconstructed resonance mass for events passing the selection cuts are shown in
Fig.~\ref{fig:kinematicswp} for the $W'$ analysis and
in Fig.~\ref{fig:kinematicskkg20} for the $KKg$ analysis.

~

Tables~\ref{tab:finalwp} and ~\ref{tab:finalkkg5} show the number of signal 
and background events passing the final selection described above for the $W'$ and $KKg$ analyses, respectively, 
at various pileup configurations, luminosities, and energies. 

\begin{table}[H]
\begin{center}
\renewcommand{\arraystretch}{1.4}
\begin{tabular}{lccc}
\hline
\multirow{2}{*}{Final Selection} & \multicolumn{2}{c}{14~TeV}  & 33~TeV  \\ 
 &  300~$fb^{-1}$&  3000~$fb^{-1}$  & 3000~$fb^{-1}$  \\ 
\hline
$W'$ (1~TeV) & 2.23E+04/1.76E+04 & 7.07E+04/3.89E+04 & 2.24E+05/4.02E+04 \\ 
$W'$ (2~TeV) & 1.07E+03/1.09E+03 & 3.50E+03/2.50E+03 & 1.95E+04/3.89E+03 \\ 
$W'$ (3~TeV) & 5.60E+01/8.46E+01 & 1.78E+02/2.19E+02 & 1.79E+03/6.92E+02 \\ 
$W'$ (4~TeV) & 4.19E+00/1.10E+01 & 1.29E+01/1.86E+01 & 2.85E+02/8.37E+01 \\ 
$W'$ (5~TeV) & 4.56E-01/2.36E+00 & 1.59E+00/4.29E+00 & 5.60E+01/3.31E+01 \\ 
$W'$ (6~TeV) & 9.69E-02/7.83E-01 & 2.85E-01/1.05E+00 & 1.48E+01/1.10E+01 \\ 
$W'$ (7~TeV) & 1.93E-02/2.20E-02 & 5.71E-02/3.92E-02 & 4.64E+00/5.00E+00 \\
$W'$ (8~TeV) &  &  & 1.79E+00/2.78E+00 \\
$W'$ (9~TeV) &  &  & 6.87E-01/6.95E-01 \\
$W'$ (10~TeV) &  &  & 3.70E-01/6.95E-01 \\
$W'$ (11~TeV) &  &  & 2.04E-01/2.73E-01 \\
$W'$ (12~TeV) &  &  & 8.55E-02/1.34E-01 \\
\hline
\end{tabular}
\caption{Signal and background event yields after the final selection 
for different integrated luminosities and CM energies for the $W'$ analysis (represented as signal/total 
background).} 
\label{tab:finalwp}
\end{center}
\end{table}

\begin{table}[H]
\begin{center}
\renewcommand{\arraystretch}{1.4}
\begin{tabular}{lccc}
\hline
\multirow{2}{*}{Final Selection} & \multicolumn{2}{c}{14~TeV}  & 33~TeV  \\ 
 &  300~$fb^{-1}$&  3000~$fb^{-1}$  & 3000~$fb^{-1}$  \\ 
\hline
$KKg$ (1~TeV) & 1.49E+04/6.41E+04 & 5.01E+04/1.71E+05 & 2.26E+05/2.14E+05 \\
$KKg$ (2~TeV) & 1.21E+03/7.34E+03 & 4.43E+03/2.06E+04 & 2.80E+04/3.28E+04 \\
$KKg$ (3~TeV) & 7.69E+01/1.13E+03 & 2.81E+02/2.67E+03 & 3.61E+03/7.12E+03 \\
$KKg$ (4~TeV) & 6.84E+00/1.98E+02 & 2.67E+01/5.72E+02 & 6.29E+02/1.95E+03 \\
$KKg$ (5~TeV) & 7.93E-01/2.33E+01 & 3.04E+00/1.07E+02 & 1.35E+02/9.91E+02 \\
$KKg$ (6~TeV) & 7.88E-02/3.51E+00 & 3.49E-01/1.32E+01 & 3.56E+01/2.40E+02 \\
$KKg$ (7~TeV) & 9.12E-03/1.51E+00 & 4.98E-02/5.17E+00 & 1.28E+01/1.48E+02 \\
$KKg$ (8~TeV) &  &  & 4.74E+00/1.30E+02 \\
$KKg$ (9~TeV) &  &  & 1.43E+00/1.22E+02 \\
$KKg$ (10~TeV) &  &  & 7.16E-01/1.44E+01 \\
$KKg$ (11~TeV) &  &  & 2.40E-01/1.15E+01 \\
$KKg$ (12~TeV) &  &  & 6.52E-02/1.09E+01 \\
\hline
\end{tabular}
\caption{Signal and background event yields after the final selection for different
integrated luminosities and CM energies for the $KKg$ analysis (represented as signal/total background).}
\label{tab:finalkkg5}
\end{center}
\end{table}

\begin{figure}[H]
  \begin{center}
    \subfigure[]{
      \includegraphics[width=0.35\textwidth]{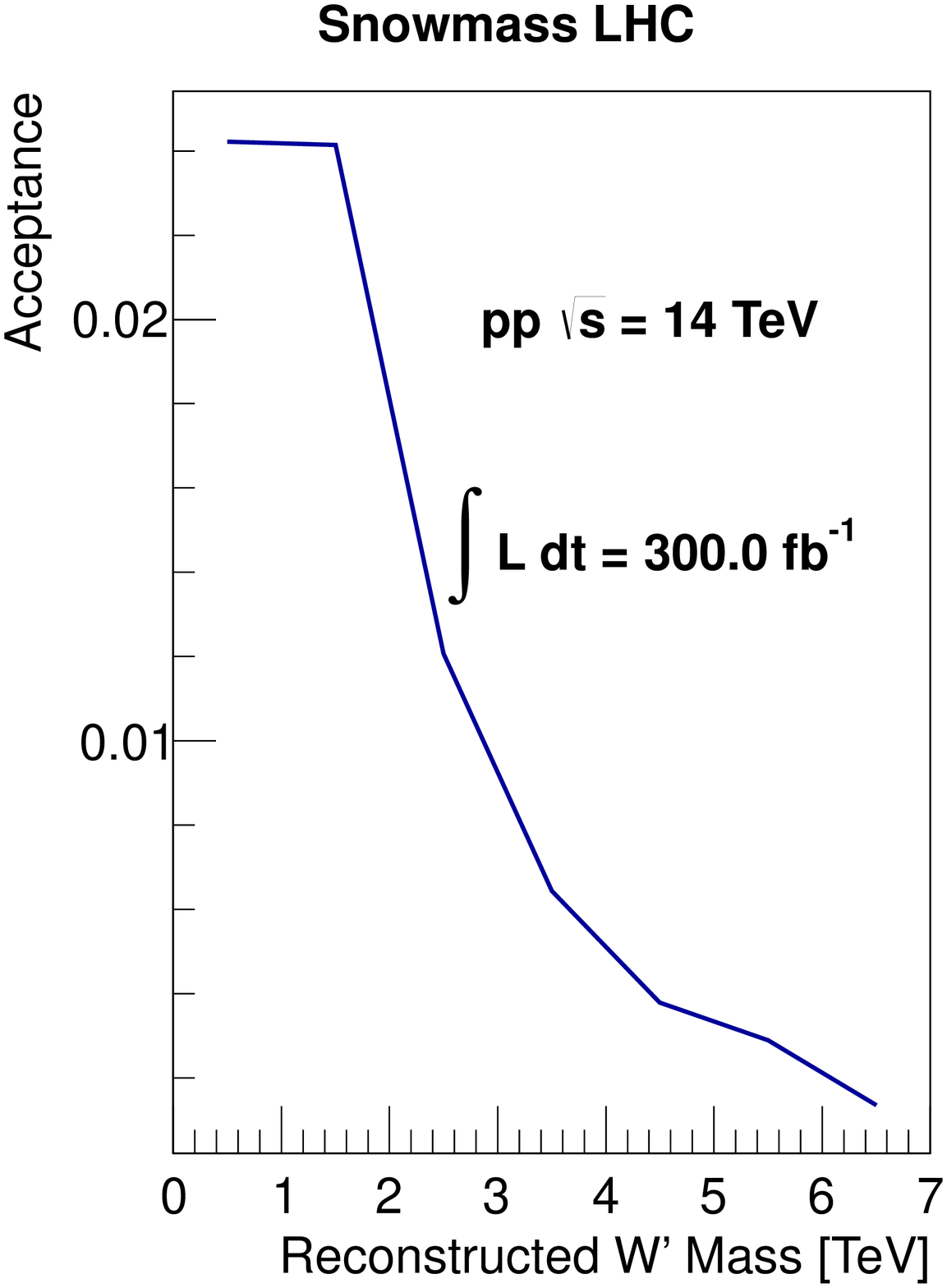}
      \label{fig:acctwa}
    }
    \subfigure[]{
      \includegraphics[width=0.35\textwidth]{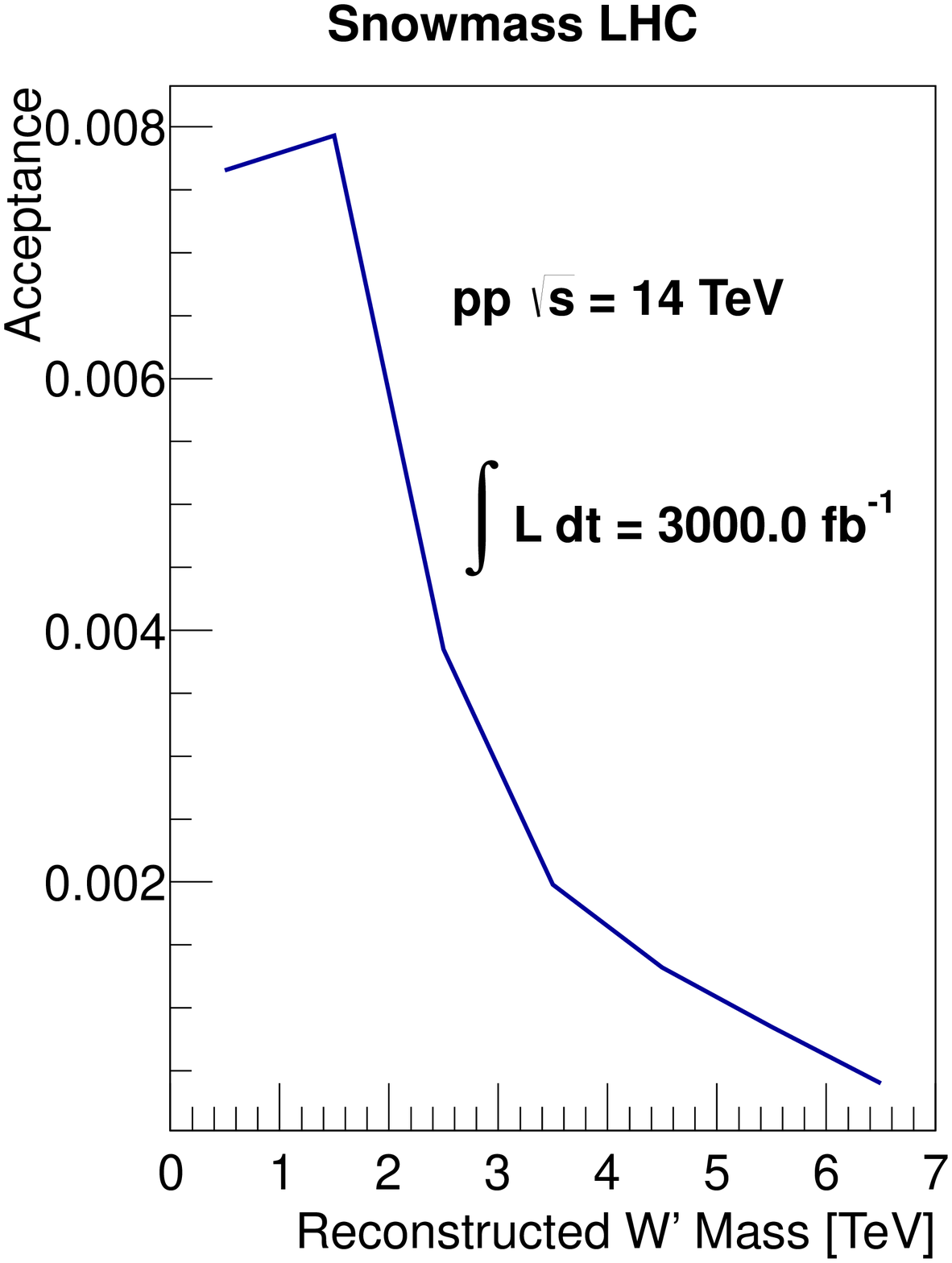}
      \label{fig:acctwb}
    }
    \subfigure[]{
      \includegraphics[width=0.35\textwidth]{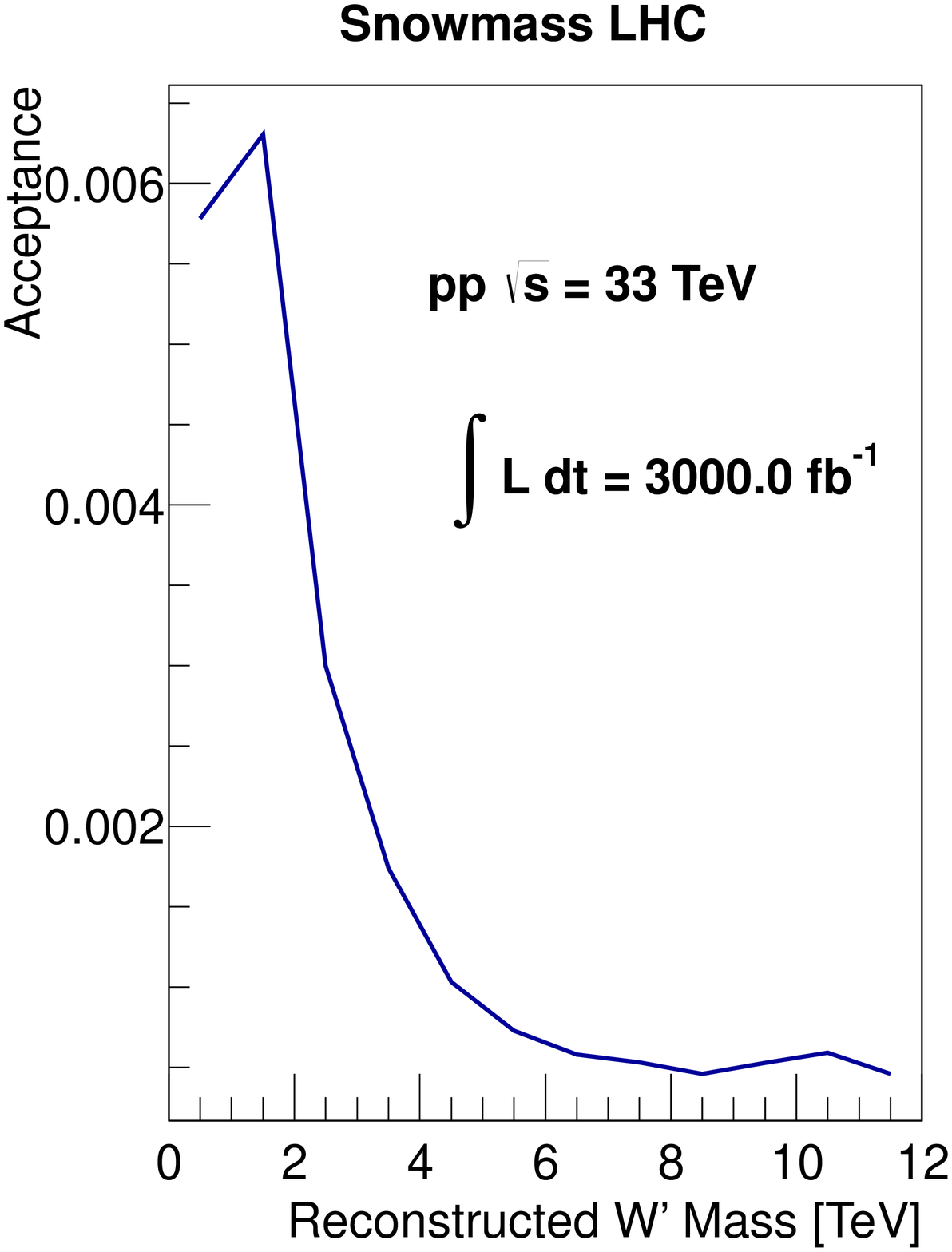}
      \label{fig:acctwc}
    }
    \caption{The acceptance of the $W'$ analysis at 14~TeV and 300~$fb^{-1}$ luminosity (a),
      14~TeV and 3000~$fb^{-1}$ luminosity (b), and 33~TeV and 3000~$fb^{-1}$ luminosity (c).}
    \label{fig:acceptancetotalwp}
  \end{center}
\end{figure}

\begin{figure}[H]
  \begin{center}
    \subfigure[]{
      \includegraphics[width=0.35\textwidth]{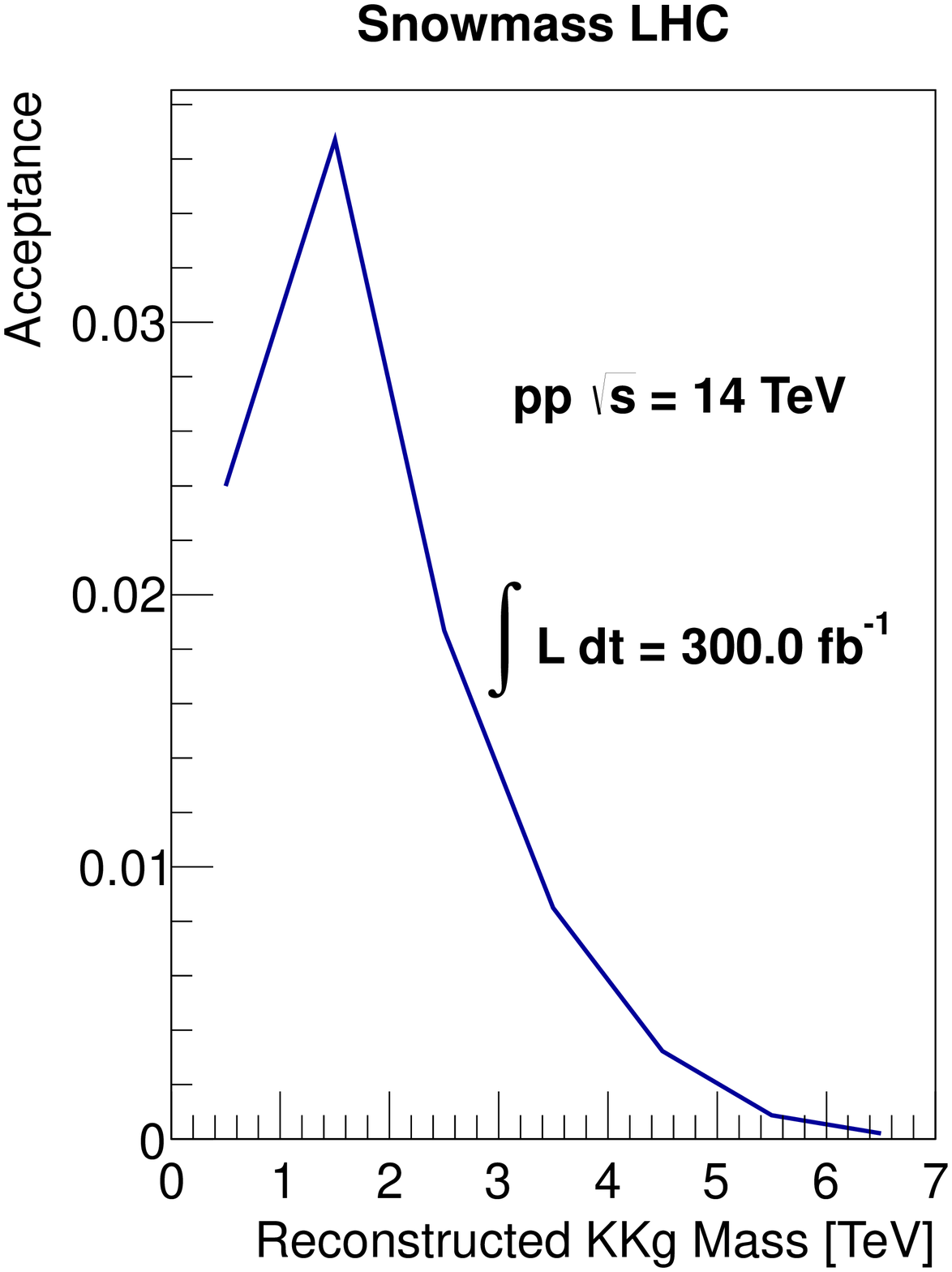}
      \label{fig:acctkkga}
    }
    \subfigure[]{
      \includegraphics[width=0.35\textwidth]{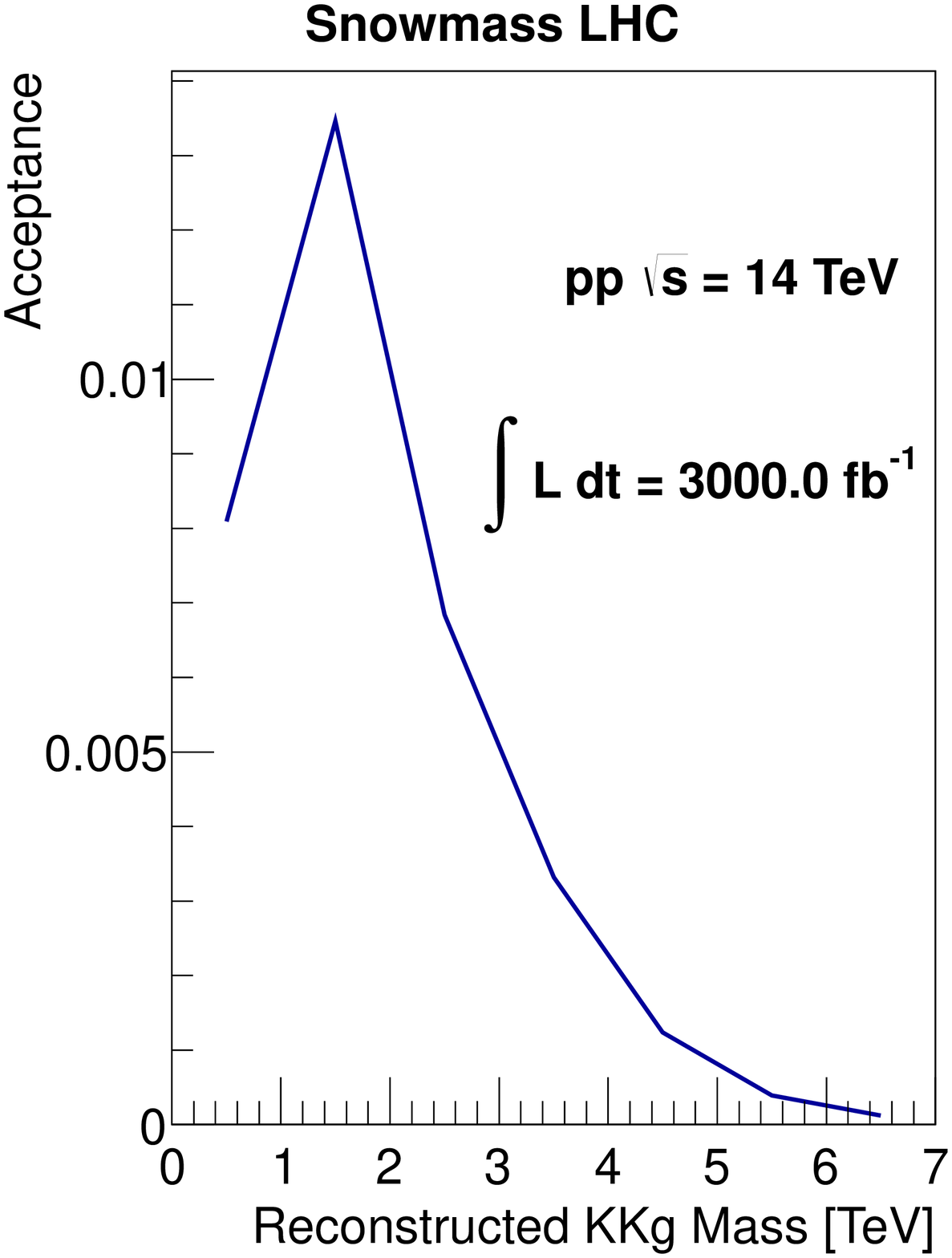}
      \label{fig:acctkkgb}
    }
    \subfigure[]{
      \includegraphics[width=0.35\textwidth]{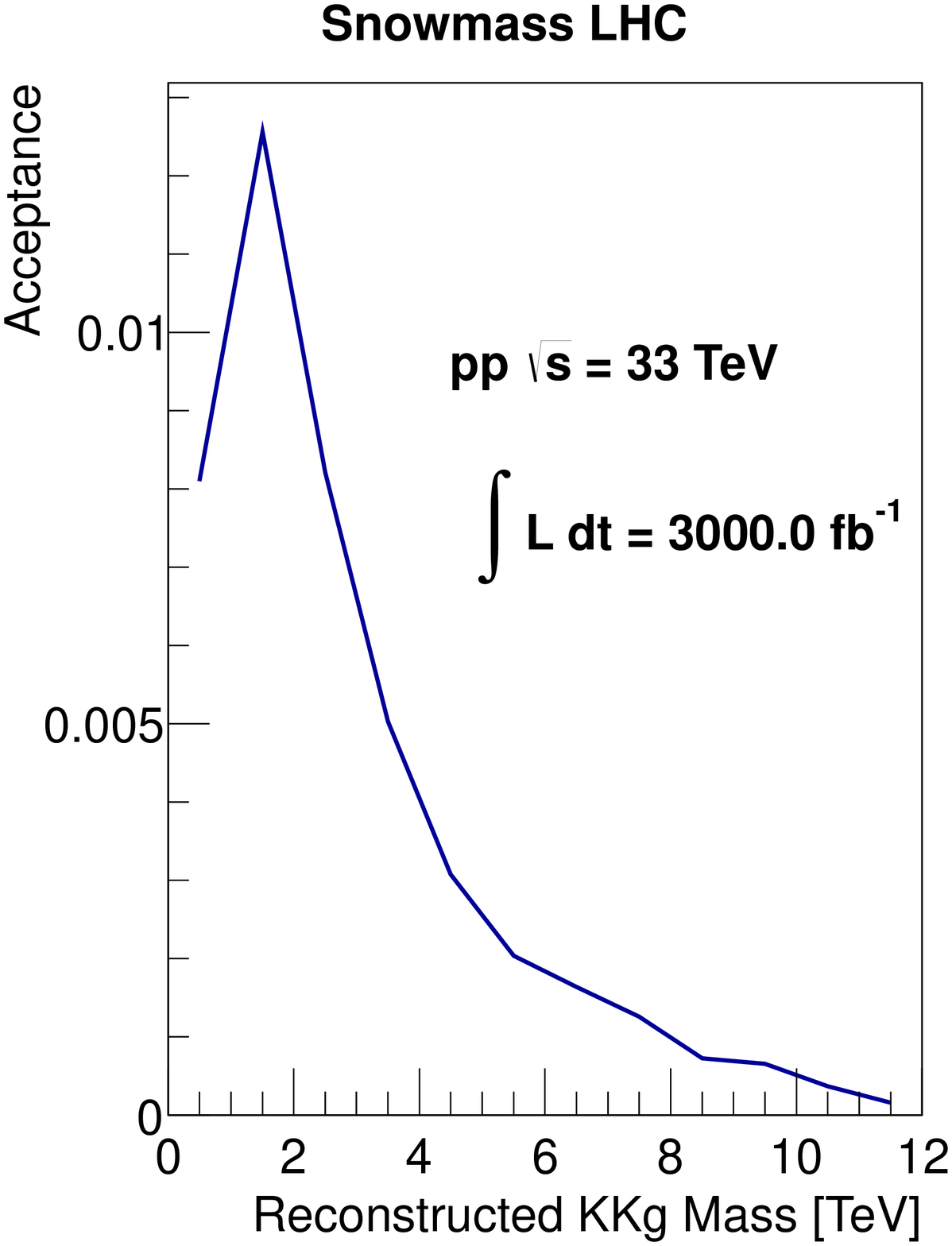}
      \label{fig:acctkkgc}
    }
    \caption{The acceptance of the $KKg$ analysis: (a)  at 14~TeV and 300~$fb^{-1}$ luminosity,
      (b) at 14~TeV and 3000~$fb^{-1}$ luminosity, (c) at 33~TeV and 3000~$fb^{-1}$ luminosity.}
    \label{fig:acceptancetotalkkg20}
  \end{center}
\end{figure}

We define the acceptance as the ratio of 
the initial number of events ($W'$ or $KKg$ events including $W$~decay to electron or muon)
and the final number of selected events. 
The acceptance for $W'$ events to pass the selection cuts as a function of the $W'$~mass is
shown in Fig.~\ref{fig:acceptancetotalwp}. The acceptance for $KKg$ events to pass the
selection cuts as a function of the $KKg$~mass is shown in Fig.~\ref{fig:acceptancetotalkkg20}.

\begin{figure}[H]
  \begin{center}
    \subfigure[]{
      \includegraphics[width=0.35\textwidth]{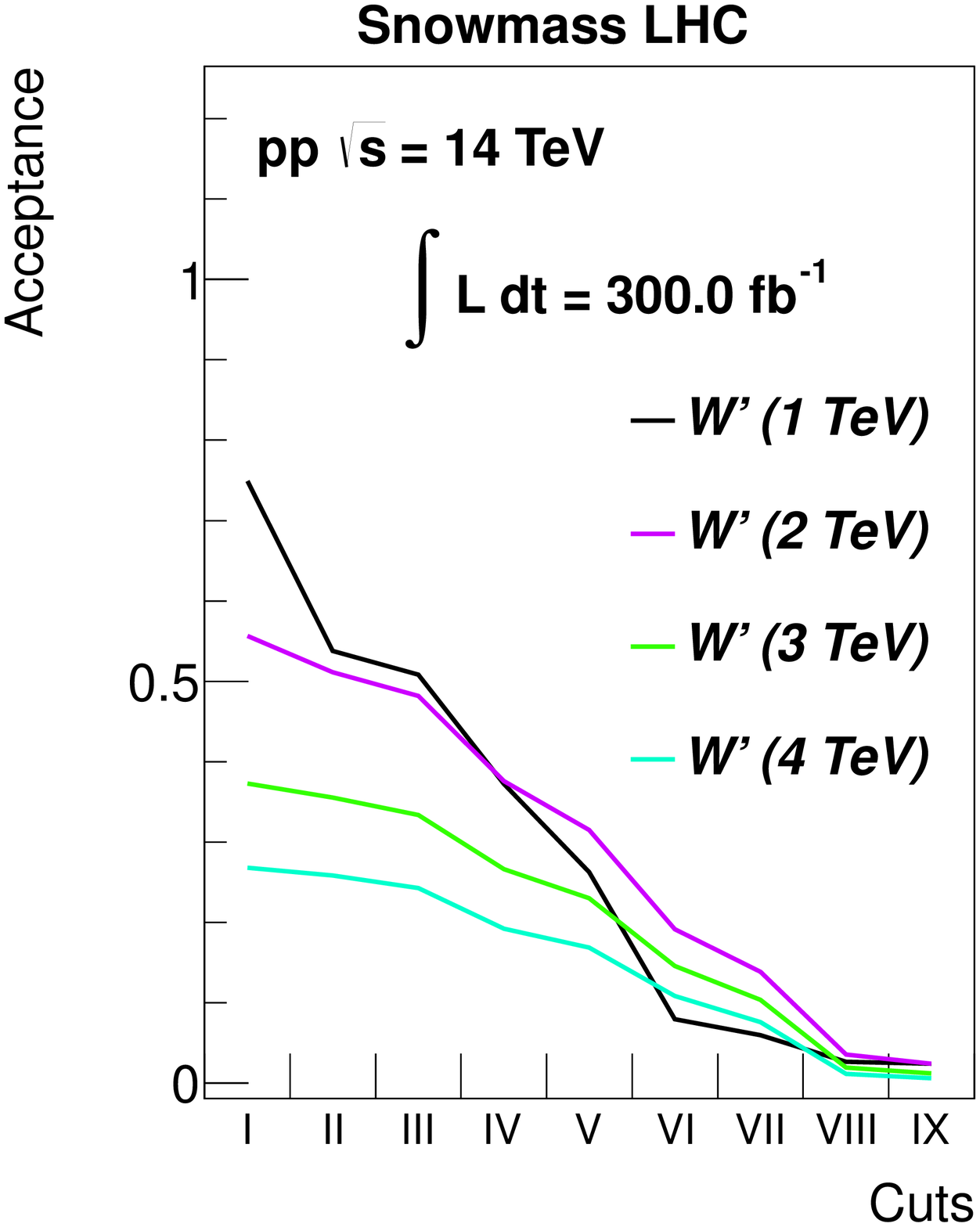}
      \label{fig:accwa}
    }
    \subfigure[]{
      \includegraphics[width=0.35\textwidth]{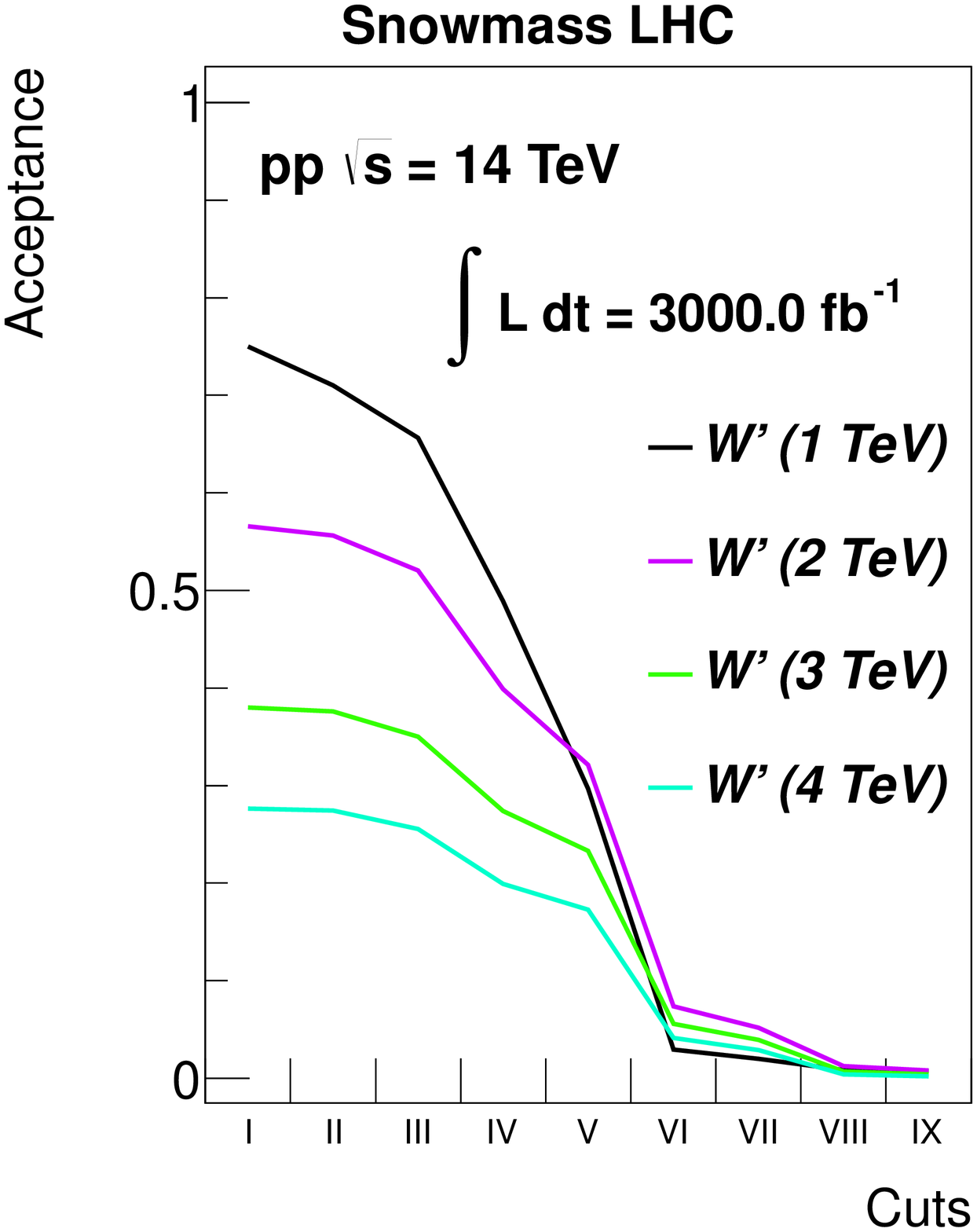}
      \label{fig:accwb}
    }
    \subfigure[]{
      \includegraphics[width=0.35\textwidth]{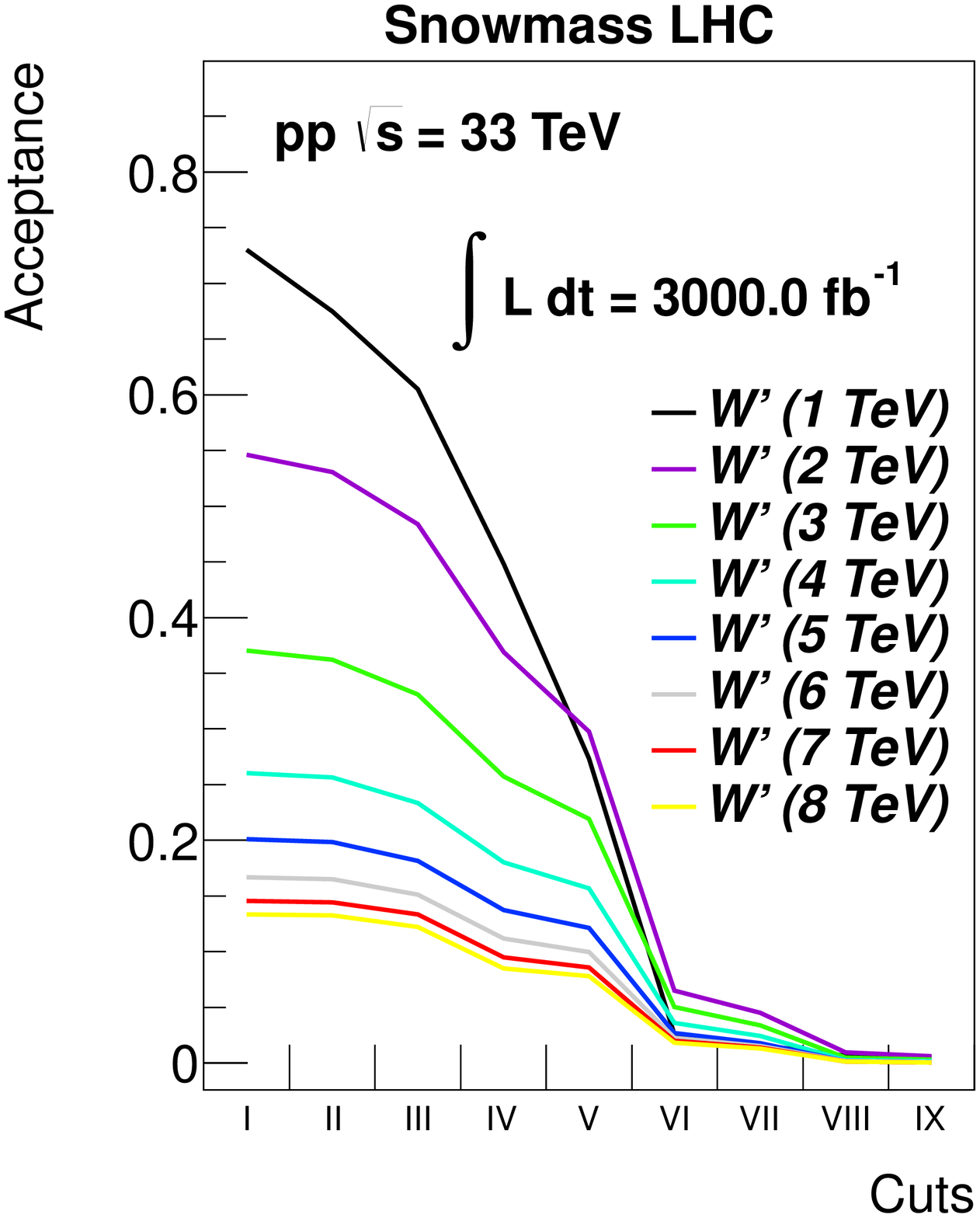}
      \label{fig:accwc}
    }
    \caption{The acceptance of the $W'$ analysis by cut:  (a) at 14~TeV and 300~$fb^{-1}$ luminosity,
      (b) at 14~TeV and 3000~$fb^{-1}$ luminosity, (c) at 33~TeV and 3000~$fb^{-1}$ luminosity. The labeling
    of the cuts is as follows: I. Number of Leptons $(=1$), II. Number of Jets ($\geq$ 2), III. Number of $b$-tagged 
    Jets ($\geq 1$), IV. $E_T^{miss}$ ($\geq$ 25~GeV), V. Lepton $p_T$ ($\geq$ 100~GeV), VI. Jet $p_T$ ($\geq$ 300~GeV), VII. Number of 
    Subjets ($\leq$ 2), VIII. Number of non-$b$-tagged Jets ($=0$), IX. $W'$ Mass Window.}
    \label{fig:acceptancecutswp}
  \end{center}
\end{figure}

\begin{figure}[H]
  \begin{center}
    \subfigure[]{
      \includegraphics[width=0.35\textwidth]{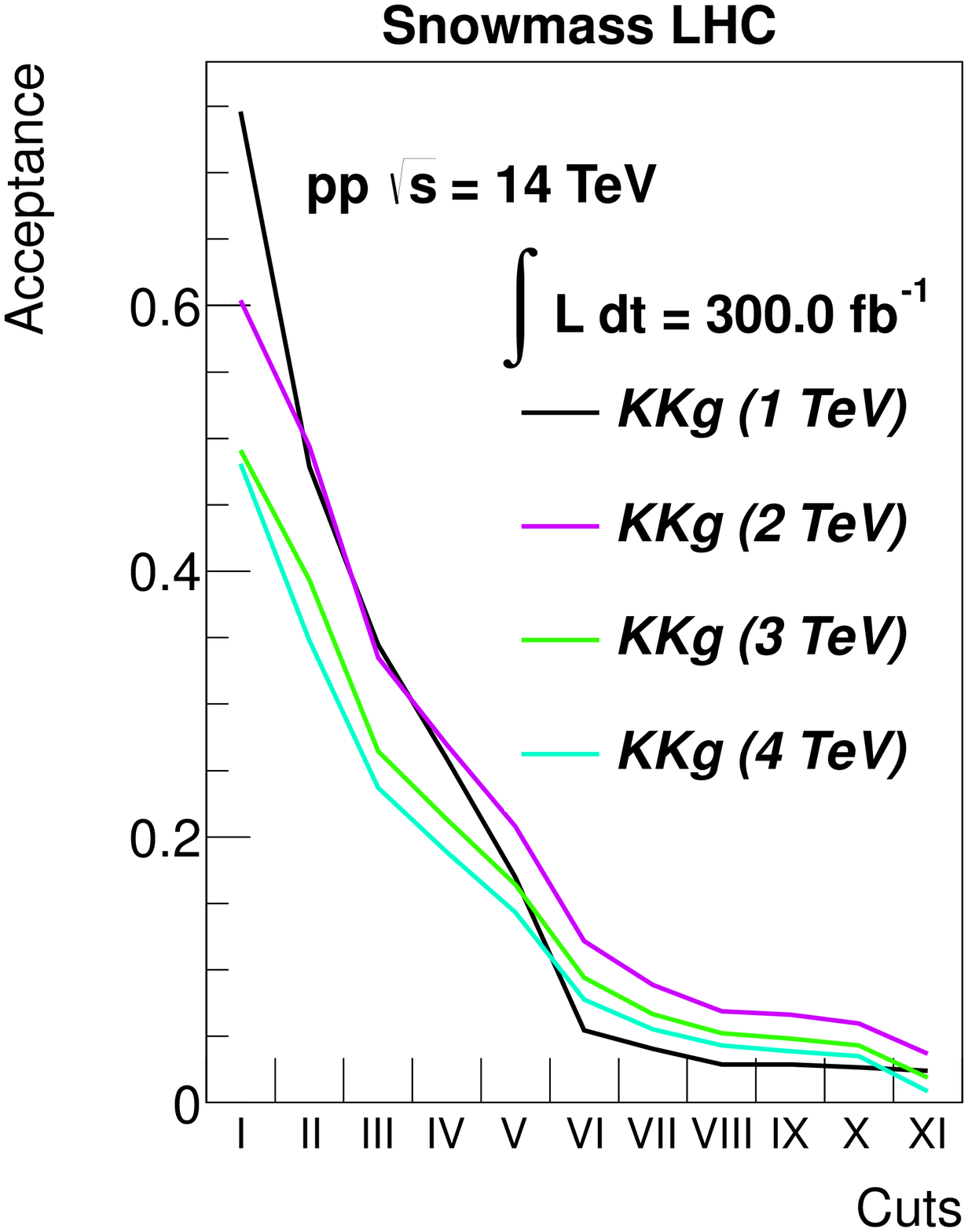}
      \label{fig:acckkg2a}
    }
    \subfigure[]{
      \includegraphics[width=0.35\textwidth]{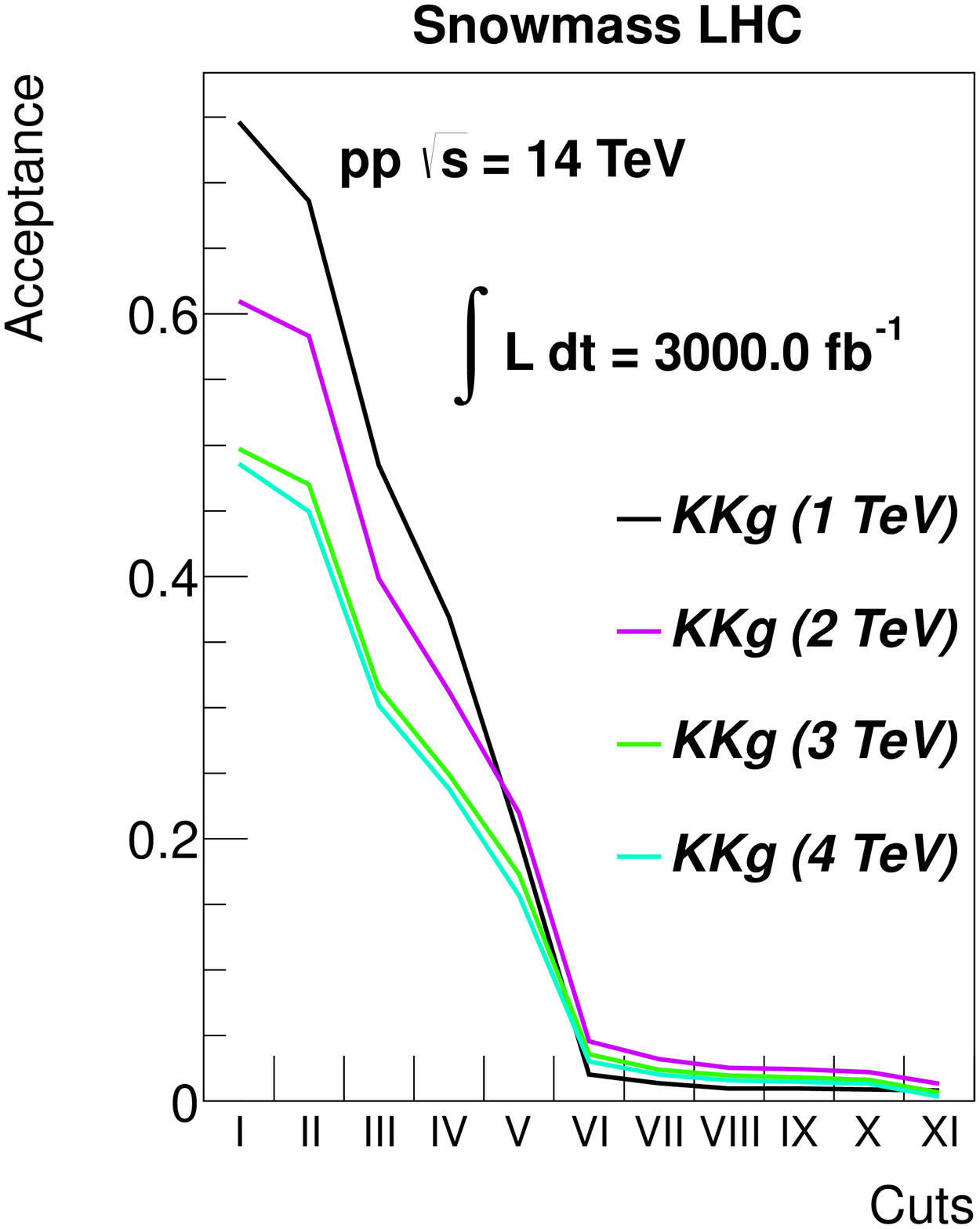}
      \label{fig:acckkg2b}
    }
    \subfigure[]{
      \includegraphics[width=0.35\textwidth]{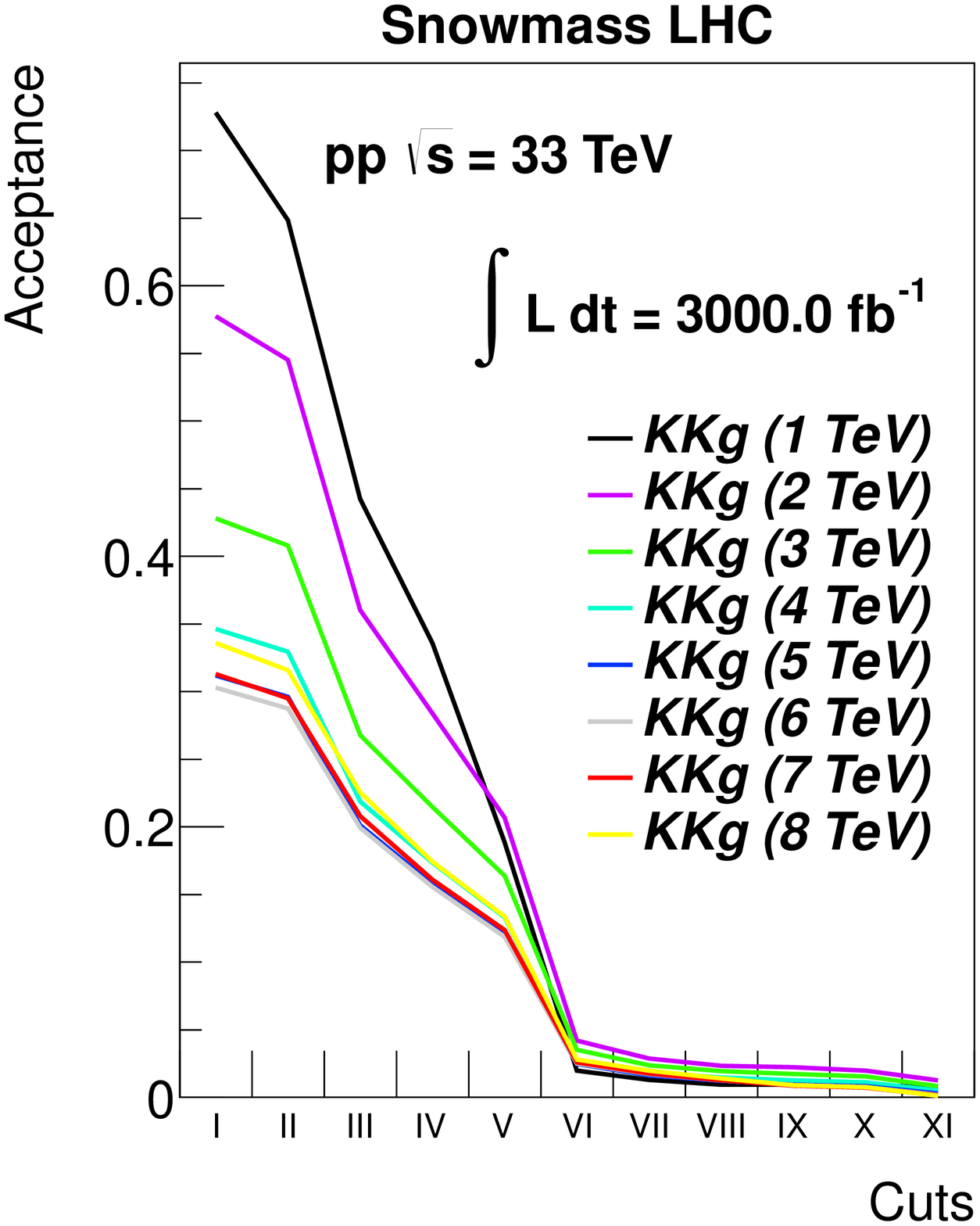}
      \label{fig:acckkg2c}
    }
    \caption{The acceptance of the $KKg$ analysis by cut: (a) at 14~TeV and 300~$fb^{-1}$ luminosity,
      (b) at 14~TeV and 3000~$fb^{-1}$ luminosity, (c) at 33~TeV and 3000~$fb^{-1}$ luminosity. The labeling
      of the cuts is as follows: I. Number of Leptons ($=1$), II. Number of Jets ($\geq 2$), III. Number of $b$-tagged 
      Jets ($\geq 1$), IV. $E_T^{miss}$ ($\geq$ 25~~GeV), V. Lepton $p_T$ ($\geq$ 100~GeV), VI. Jet $p_T$ ($\geq$ 300~GeV), 
      VII. Number of Subjets ($\leq 2$), VIII. Number of $b$-tagged Jets ($=1$), IX. Number of non-$b$-tagged 
      Jets ($=1$), X. Reconstructed top Mass ($\leq$ 200~GeV), XI. $KKg$ Mass Window.}
    \label{fig:acceptancecutskkg}
  \end{center}
\end{figure}

It is also interesting to view the acceptance as a function of the cut made, as for the $W'$ analysis 
in Fig.~\ref{fig:acceptancecutswp} and for the $KKg$ analysis in 
Fig.~\ref{fig:acceptancecutskkg}. 

\subsection{{\boldmath $W'$} Cross-Section Limits}
\label{sec:WprimeLimits}
Finally, 95$\%$ confidence limits on the $W'$ and KKg cross-sections are set assuming a
background normalization uncertainty of 10$\%$. The cross-section limits are converted into
mass limits using the theoretical dependence on the resonance mass.

\begin{figure}[H]
  \begin{center}
    \subfigure[]{
      \includegraphics[width=0.45\textwidth]{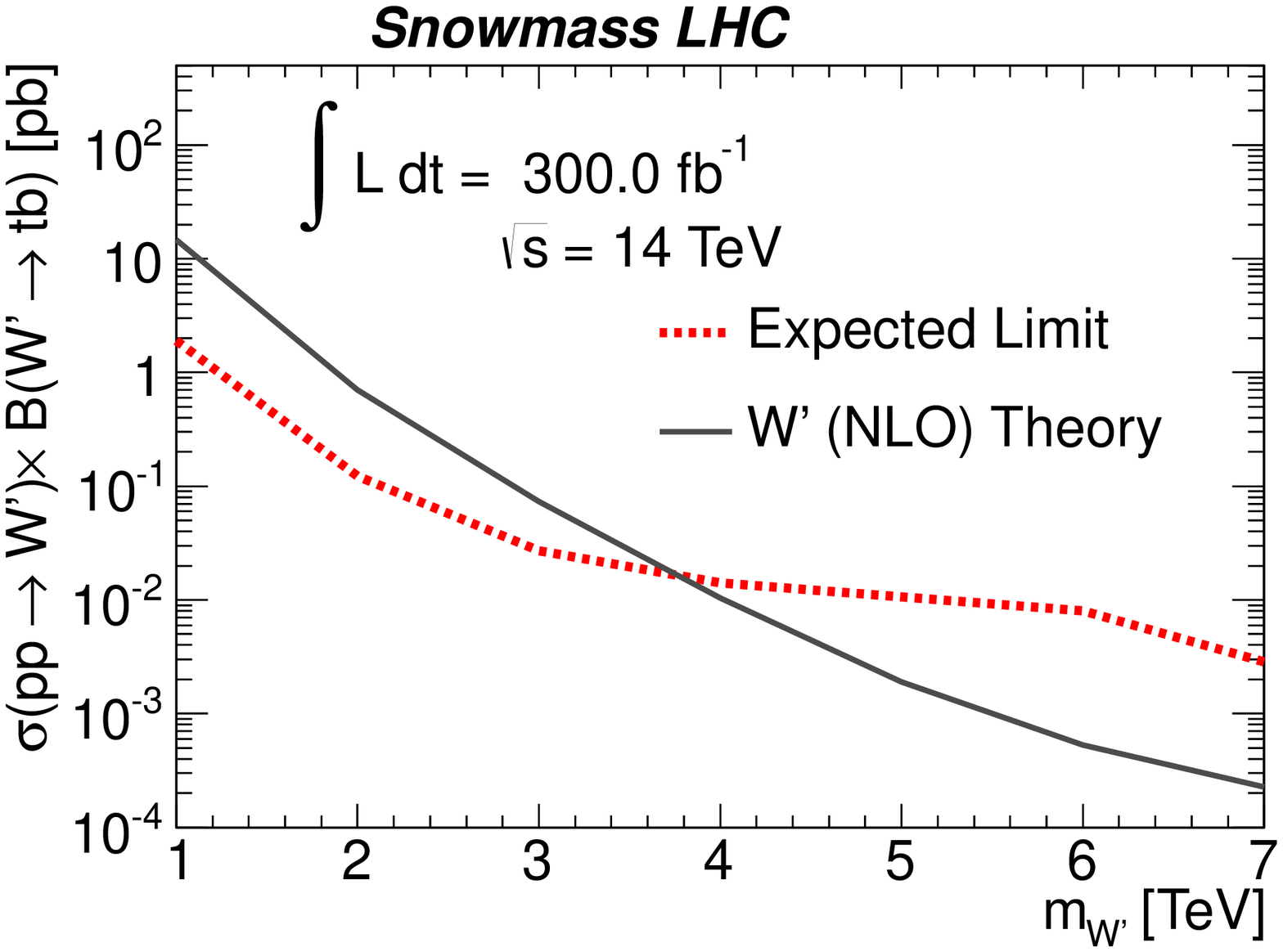}
      \label{fig:limwa}
    }
    \subfigure[]{
      \includegraphics[width=0.45\textwidth]{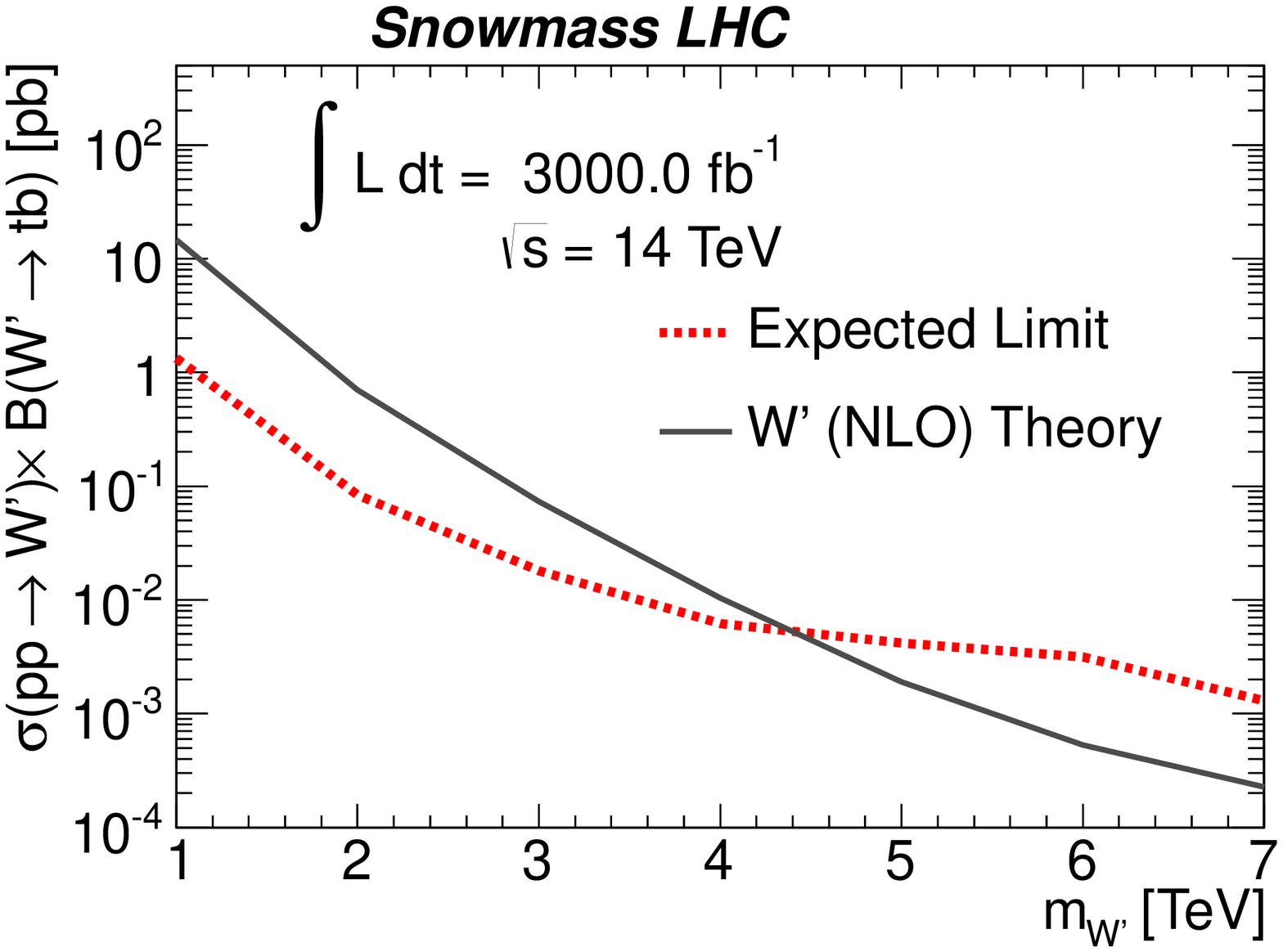}
      \label{fig:limwb}
    }
    \subfigure[]{
      \includegraphics[width=0.45\textwidth]{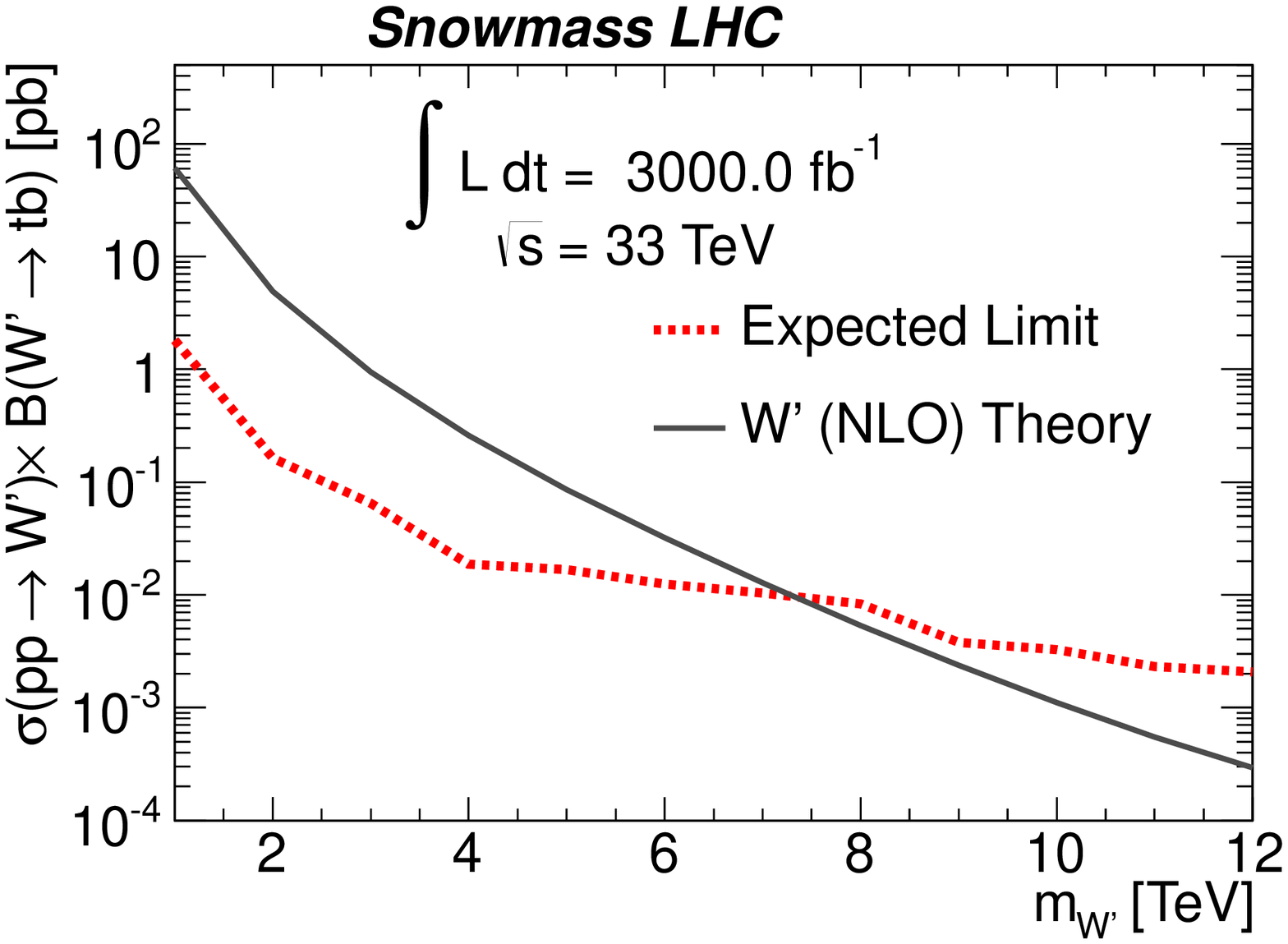}
      \label{fig:limwc}
    }
    \caption{The cross-section times branching ratio as a function of the $W'$ mass: (a) at 14~TeV and 
      300~$fb^{-1}$ luminosity, (b) at 14~TeV and 3000~$fb^{-1}$ luminosity, (c) at 33~TeV and 3000~$fb^{-1}$ 
      luminosity.}
    \label{fig:wprimelimits}
  \end{center}
\end{figure}

Fig.~\ref{fig:wprimelimits} shows the cross-section times branching ratio as a function of the $W'$ mass 
at 14~TeV and 33~TeV for different luminosities.

The $W'$  mass limit at 14~TeV is 3.8~TeV and 4.4~TeV for 300~$fb^{-1}$ and 3000~$fb^{-1}$ luminosity, respectively. 
The mass limit at 33~TeV and 3000~$fb^{-1}$ luminosity is 7.3~TeV. The expected $W'$ cross-sections are show at 
Table~\ref{tab:Wprimelimit}. 
\begin{table}[H]                                                                       
  \begin{center}                                                                         
    \renewcommand{\arraystretch}{1.4}                                                      
    \begin{tabular}{lccc}                                                        
      \hline                                                                                 
      \multirow{2}{*}{$W'$ Mass [TeV]} & \multicolumn{2}{c}{14~TeV [pb]} & 33~TeV [pb] \\           
                      & 300~$fb^{-1}$ & 3000~$fb^{-1}$& 3000~$fb^{-1}$  \\           
      \hline                                                                                 
      1~TeV &  1.9E+00 & 1.3E+00  & 1.8E+00\\
      2~TeV &  1.2E-01 & 8.4E-02  & 1.6E-01    \\  
      3~TeV &  2.7E-02 & 1.8E-02  & 6.4E-02 \\
      4~TeV &  1.4E-02 & 6.2E-03  & 1.9E-02  \\
      5~TeV &  1.1E-02 & 4.2E-03  & 1.7E-02 \\    
      6~TeV &  8.0E-03 & 3.2E-03  & 1.3E-02  \\
      7~TeV &  2.9E-03 & 1.3E-03  & 1.0E-02 \\
      8~TeV &          &          & 8.3E-03      \\
      9~TeV &          &          & 3.8E-03      \\
      10~TeV &         &          & 3.3E-03      \\
      11~TeV &         &          & 2.3E-03      \\
      12~TeV &         &          & 2.1E-03      \\
      \hline                                                                                 
    \end{tabular}                                                                          
    \caption{Expected cross-section limits for various $W'$ masses at 14~TeV and 33~TeV.}
    \label{tab:Wprimelimit}
  \end{center}
\end{table}

\subsection{KKg Cross-Section Limits}
\label{sec:KKgLimits}
\begin{figure}[H]
  \begin{center}
    \subfigure[]{
      \includegraphics[width=0.45\textwidth]{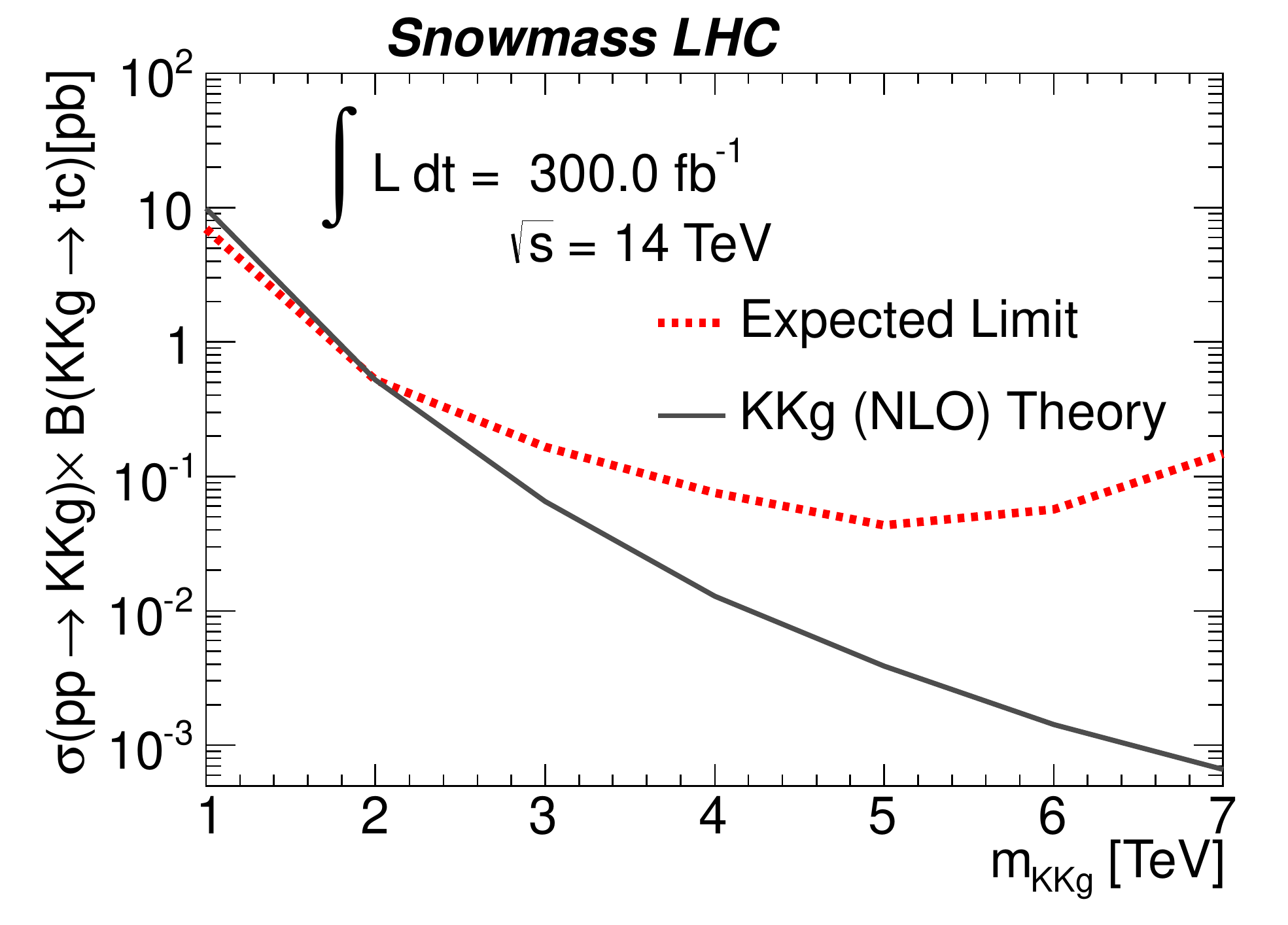}
      \label{fig:limkkga}
    }
    \subfigure[]{
      \includegraphics[width=0.45\textwidth]{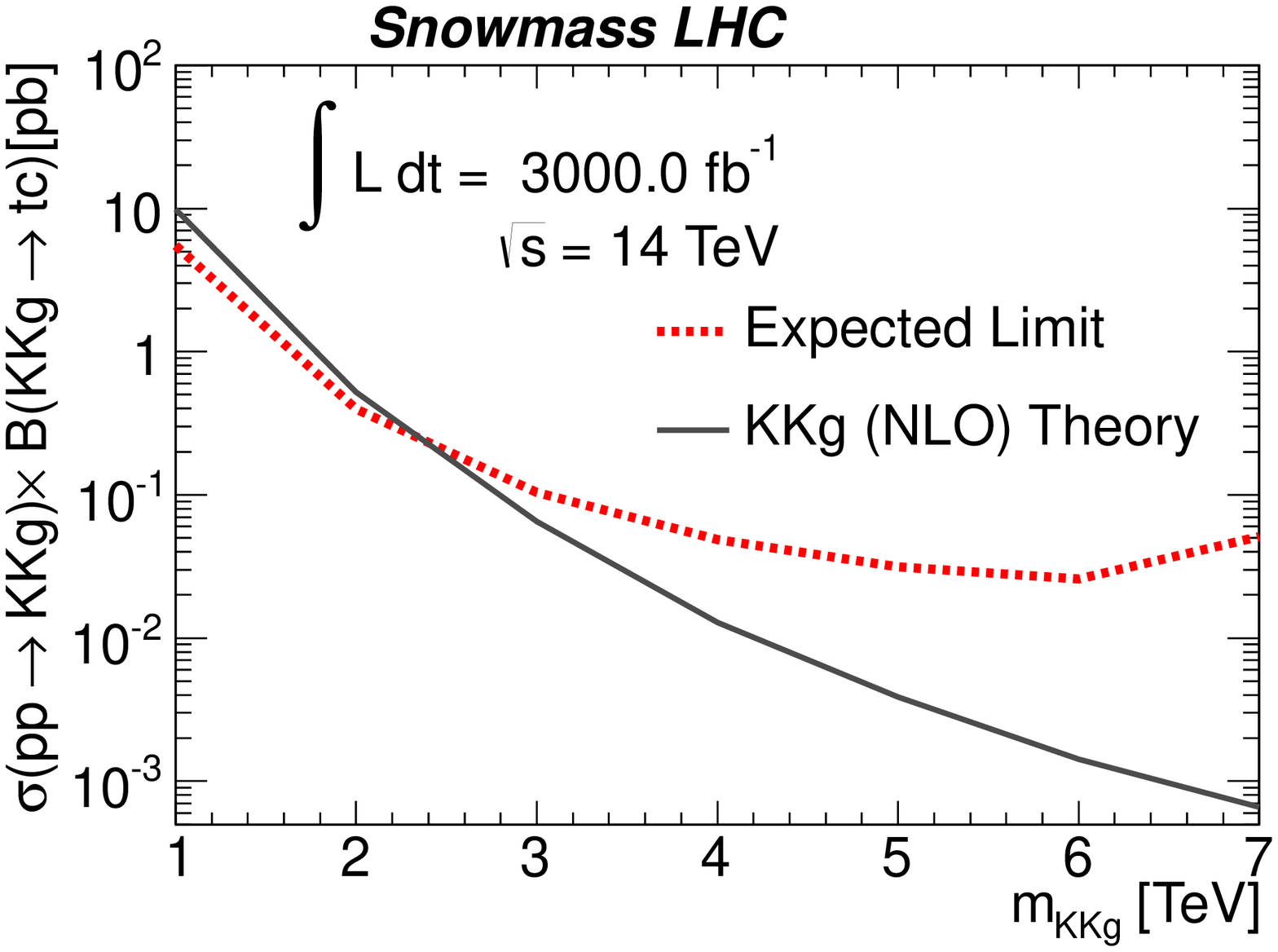}
      \label{fig:limkkgb}
    }
    \subfigure[]{
      \includegraphics[width=0.45\textwidth]{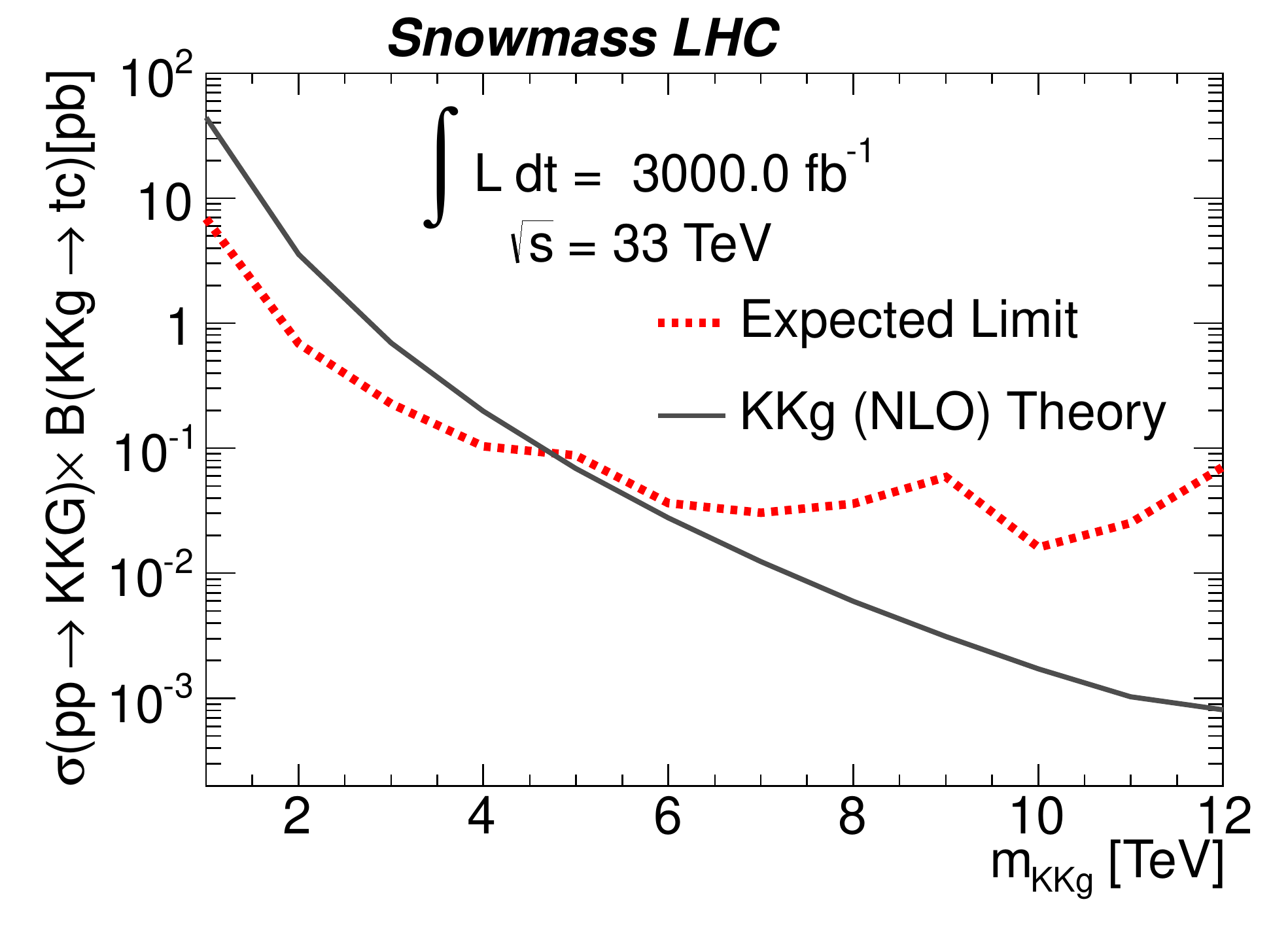}
      \label{fig:limkkgc}
    }
    \caption{The cross-section times branching ratio as a function of the $KKg$ mass: (a) at 14~TeV and 
      300~$fb^{-1}$ luminosity, (b) at 14~TeV and 3000~$fb^{-1}$ luminosity, (c) at 33~TeV and 3000~$fb^{-1}$ 
      luminosity.}
    \label{fig:KKglimits}
  \end{center}
\end{figure}

Fig.~\ref{fig:KKglimits} shows the cross-section times branching ratio as a function of the $KKg$ mass 
at 14~TeV and 33~TeV for different luminosities for the $KKg$ analysis.

The $KKg$  mass limit for at 14~TeV and 300~$fb^{-1}$ luminosity is 2.0~TeV, at 14~TeV and 3000~$fb^{-1}$ 
luminosity is 2.4~TeV, and 33~TeV and 3000~$fb^{-1}$ luminosity is 4.7~TeV. The expected $KKg$ cross 
sections for the $KKg$ analysis are show at Table~\ref{tab:KKglimit}. 
 \begin{table}[H]                                                                       
   \begin{center}                                                                         
     \renewcommand{\arraystretch}{1.4}                                                      
     \begin{tabular}{lccc}                                                        
       \hline                                                                                 
       $KKg$ Mass [TeV] & \multicolumn{2}{c}{14~TeV [pb]} & 33~TeV [pb] \\           
       & 300~$fb^{-1}$ & 3000~$fb^{-1}$& 3000~$fb^{-1}$  \\           
       \hline                                                                                 
       1~TeV  & 7.0E+00 & 5.5E+00  &6.9E+00  \\ 
       2~TeV  & 5.2E-01 & 4.0E-01 & 6.8E-01 \\ 
       3~TeV  & 1.7E-01 & 1.0E-01 & 2.3E-01 \\ 
       4~TeV  & 7.5E-02 & 4.9E-02 & 1.0E-01 \\ 
       5~TeV  & 4.3E-02 & 3.1E-02 & 8.8E-02 \\ 
       6~TeV  & 5.7E-02 & 2.6E-02 & 3.7E-02 \\ 
       7~TeV  & 1.5E-01 & 5.1E-02 & 3.0E-02 \\ 
       8~TeV  &  &  &             3.6E-02\\
       9~TeV  &  &  &             5.6E-02\\
       10~TeV &  &  &             1.6E-02\\
       11~TeV &  &  &             2.5E-02\\
       12~TeV &  &  &             7.1E-02\\
       \hline                                                                                 
     \end{tabular}                                                                          
     \caption{Expected cross-section limits for various $KKg$ masses at 14~TeV and 33~TeV for the 
       $KKg$ analysis.}
     \label{tab:KKglimit}
   \end{center}
 \end{table}

%

\section{Conclusions}
\label{sec:conclusions}

We have presented the possibilities for future searches for beyond the standard model
particles at the LHC. The 14~TeV LHC is sensitive to $W'$~bosons decaying to $tb$ with
right-handed couplings masses up to 3.8~TeV with 300~fb$^{-1}$ and up to 4.0~TeV 
with 300~fb$^{-1}$. A 33~TeV collider can reach $W'$ masses up to 7~TeV. 
This will greatly improve the current limit of about 2~TeV set by 
ATLAS~\cite{Aad:2012ej, WprimeCONF} and CMS~\cite{CMSWprime}. 

The sensitivity for FCNC $KKg$ is not as good because the $KKg\rightarrow tc$ final state
only has one $b$~quark, leading to large backgrounds. Only for a $KKg \rightarrow tc$ cross
section as 
large as 40$\%$ of the $KKg \rightarrow t\bar{t}$ NLO cross-section it will be possible to set
limits, and even then only for $KKg$ masses up to 2~TeV at the 14~TeV LHC. 
The situation is better at a 33~TeV hadron collider which has sensitivity up to 5~TeV for 
a $KKg \rightarrow tc$ cross-section that is 40$\%$ of the $KKg \rightarrow t\bar{t}$ NLO
cross-section. At low $KKg$ masses, a 33~TeV hadron collider has sensitivity to
$KKg\rightarrow tc$ with a cross-section of about 5\% of the $KKg \rightarrow t\bar{t}$ NLO cross-section.

\begin{acknowledgments}
This work was supported in part by the U.S. National Science Foundation under Grants No. PHY-0952729 and PHY-1068318.
\end{acknowledgments}

\bibliography{tbres}

\end{document}